\documentclass[journal]{IEEEtran}
\usepackage{cite}
\usepackage{color}
\usepackage{amsmath,amssymb,amsfonts}
\usepackage{algorithmic}
\usepackage{graphicx}
\usepackage{textcomp}

\usepackage{multirow}
\newlength\savewidth
\newcommand\shline{\noalign{\global\savewidth\arrayrulewidth
                            \global\arrayrulewidth 0.8pt}
                   \hline
                   \noalign{\global\arrayrulewidth\savewidth}}  
\usepackage{bbding}
\usepackage{pifont}
\usepackage{wasysym}

\usepackage{amssymb}
\usepackage[export]{adjustbox}
\usepackage{makecell}
\usepackage{url}

\begin{document}
\title{PUERT: Probabilistic Under-sampling and Explicable Reconstruction Network for CS-MRI}
\author{Jingfen~Xie, Jian~Zhang, Yongbing~Zhang, Xiangyang~Ji

% \thanks{Manuscript received October 14, 2021, revised March 23, 2022 and accepted April 11, 2022. This work was supported in part by the Ministry of Science and Technology of China (Grant No. 2020AA0108202) and National Natural Science Foundation of China (61902009). \textit{(Corresponding author: Jian Zhang)} }
\thanks{J.~Xie and J.~Zhang are with the School of Electronic and Computer Engineering, Peking University Shenzhen Graduate School, Shenzhen 518055, China. J.~Zhang is also with the Peng Cheng Laboratory, Shenzhen 518055
China. (E-mail: xiejf@stu.pku.edu.cn; zhangjian.sz@pku.edu.cn)
}
\thanks{Y.~Zhang is with the the School of Computer Science and Technology, Harbin Institute of Technology (Shenzhen), Shenzhen 518055, China. (E-mail: ybzhang08@hit.edu.cn)
}
\thanks{X.~Ji is with the Department of Automation, Tsinghua University, Beijing 100084, China. (E-mail: xyji@tsinghua.edu.cn)
}
}

\markboth{IEEE Journal of Selected Topics in Signal Processing, ~2022}%
{Shell \MakeLowercase{\textit{et al.}}: Bare Demo of IEEEtran.cls for IEEE Journals}

\maketitle

\begin{abstract}
Compressed Sensing MRI (CS-MRI) aims at reconstructing de-aliased images from sub-Nyquist sampling $k$-space data to accelerate MR Imaging, thus presenting two basic issues, \textit{i.e.}, where to sample and how to reconstruct. To deal with both problems simultaneously, we propose a novel end-to-end Probabilistic Under-sampling and Explicable Reconstruction neTwork, dubbed PUERT, to jointly optimize the sampling pattern and the reconstruction network. Instead of learning a deterministic mask, the proposed sampling subnet explores an optimal probabilistic sub-sampling pattern, which describes independent Bernoulli random variables at each possible sampling point, thus retaining robustness and stochastics for a more reliable CS reconstruction. A dynamic gradient estimation strategy is further introduced to gradually approximate the binarization function in backward propagation, which efficiently preserves the gradient information and further improves the reconstruction quality. Moreover, in our reconstruction subnet, we adopt a model-based network design scheme with high efficiency and interpretability, which is shown to assist in further exploitation for the sampling subnet.
Extensive experiments on two widely used MRI datasets demonstrate that our proposed PUERT not only achieves state-of-the-art results in terms of both quantitative metrics and visual quality but also yields a sub-sampling pattern and a reconstruction model that are both customized to training data. \footnote{\color{blue} For reproducible research, the source codes and training models of our PUERT are available at \url{https://github.com/chuan1093/PUERT}.}
\end{abstract}


\section{Introduction}
\IEEEPARstart{M}{agnetic} Resonance Imaging (MRI) is a widely-used biomedical imaging technology that enjoys superior benefits of good soft-tissue contrast, non-ionizing radiation, and the availability of multiple tissue contrasts. A main challenge lies in how to reduce the long scan time so as to improve accessibility and decrease costs. One solution is to accelerate MRI via Compressed Sensing (CS)  \cite{lustig2008compressed,gamper2008compressed}. 
In Compressed Sensing MRI (CS-MRI), sub-Nyquist sampling \cite{lustig2008compressed} is utilized to get under-sampled $k$-space data, \textit{i.e.}, part of the Fourier transform of the image, following a predetermined sampling mask. Then, given a sub-sampled set of measurements, a CS-MRI reconstruction algorithm with high quality and a fast speed is expected to reconstruct the full-resolution MRI without aliased artifacts. Thus, there exist two important issues within CS-MRI, \textit{i.e.}, where to sample and how to reconstruct.

For sampling schemes, some popular patterns in CS-MRI include Cartesian \cite{haldar2010compressed} with skipped lines, Random Uniform \cite{gamper2008compressed} and Variable Density (VD) \cite{wang2009variable}, thanks to their simplicity and good performance when coupled with reconstruction methods. 
These sampling schemes mostly follow variable-density probability density functions, based on the empirical observation that lower frequencies should be sampled more densely than high frequencies to promote image recovery.
Another common Poisson disc sampling strategy \cite{levine20173d} adopts sampling locations separated by a minimum distance \cite{lustig2010spirit} in addition to following a density, thus further exploiting redundancies in parallel MRI. However, these masks are designed heuristically and independently, lacking the ability to adapt to specific data and recovery methods, leaving much room for further improvement.

For MRI reconstruction, there exist various works improving it from many aspects. On one hand, traditional model-based MRI restoration has been widely studied \cite{lustig2008compressed,haldar2010compressed,otazo2010combination,lustig2007sparse,qu2012undersampled,yang2010fast,block2007undersampled,trzasko2008highly,liang2009accelerating,zhao2018cream}. These methods usually adopt iterative optimization, resulting in over-smoothed recovery and long consuming time. 
On the other hand, data-driven methods have been introduced as a promising alternative \cite{xu2014deep,hyun2018deep,ronneberger2015u,schlemper2017deep,zheng2019cascaded,sun2018compressed,liu2020rare,souza2020enhanced,chen2022ai} with high quality and a fast speed. In \cite{hyun2018deep}, a widely-used convolutional neural network called U-Net \cite{ronneberger2015u} is used to reconstruct MR images. In \cite{schlemper2017deep}, a cascaded CNN with a data consistency layer is presented to further ensure measurement fidelity. 
Most recently, some Deep Unfolding Networks (DUNs) \cite{zhang2018ista,yang2016deep,aggarwal2018modl,liu2020deep,qin2018convolutional,hosseini2020dense,song2021memory} are developed to integrate the interpretability of traditional model-based approaches and the efficiency of data-driven methods, thus yielding a better recovery performance.
Note that DUNs are not limited to CS-MRI, but also widely studied in CS reconstruction \cite{zhang2020optimization,you2021coast,you2021ista,wu2021dense,wu2021spatial}.
As an instance, a state-of-the-art method ISTA-Net \cite{zhang2018ista} unfolds the traditional iterative shrinkage-thresholding algorithm (ISTA) update steps to a network with a fixed number of stages, each corresponding to one iteration in ISTA. Besides, recent works have also focused on exploring different training objectives such as adversarial loss \cite{NIPS2014_5ca3e9b1,isola2017image,yang2017dagan} and cyclic loss \cite{quan2018compressed,li2021high} to enhance perceptual recovery quality. 

The above-mentioned works regard sub-sampling and reconstruction as two independent problems, in consideration of generality and simplicity. However, they ignore the fact that, in general, the optimal under-sampling pattern depends on the specific MRI anatomy and reconstruction method, thus requiring more customization. 

In this paper, we jointly deal with the above two issues, \textit{i.e.}, under-sampling and reconstruction, and propose a novel end-to-end Probabilistic Under-sampling and Explicable Reconstruction neTwork, dubbed PUERT and pronounced like Pu’er Tea, to achieve an efficient combination of sub-sampling learning and reconstruction network training. Specifically, considering the stochastic strategies of compressed sensing, we develop a sampling subnet that explores an optimal Bernoulli probabilistic sampling pattern instead of a deterministic mask, thus achieving robustness and stochastics for a more reliable reconstruction. Then, a dynamic gradient estimation strategy is proposed to gradually approximate the binarization function in backward propagation, which efficiently retains the gradient information and further improves the recovery quality. Moreover, we adopt an efficient reconstruction subnet unfolded by the traditional ISTA algorithm, and also emphasize the superiority of DUN in facilitating exploration and efficient training of the sampling subnet.

Overall, the main contributions of this paper are four-fold: 
\begin{itemize}
\item A novel end-to-end Probabilistic Under-sampling and Explicable Reconstruction neTwork, dubbed PUERT, for CS-MRI is proposed, which implements an efficient combination of sampling mask learning and reconstruction network training.
\item The sampling subnet explores an optimal probabilistic sampling pattern to retain robustness and stochastics, and introduces a dynamic gradient estimation strategy to enable efficient training and promote network performance.
\item The end-to-end reconstruction subnet adopts an explicable ISTA-unfolding network with high reconstruction quality and fast speed, which is also shown to facilitate further exploration for the sampling subnet.
\item Experiments show that our proposed PUERT not only performs favorably against state-of-the-arts in terms of both quantitative metrics and visual quality but also yields a sub-sampling pattern and a reconstruction model that are both customized to training data.
\end{itemize}

\section{Related Work}
Existing sampling pattern optimization schemes consist of two classes: 1) active algorithms \cite{zhang2019reducing,jin2019self} using a sampling subnet to predict the \textit{next} $k$-space sample location based on current recovery from reconstruction subnet, and 2) non-active methods directly predicting the \textit{whole} mask in every epoch. Here, we focus on the latter kind and further divide them into nested optimization methods and end-to-end learning methods.

\subsection{Nested Optimization Methods}
Nested optimization methods formulate the sampling matrix learning problem as two nested optimization issues. The outer issue is to find an optimal sub-sampling mask or trajectory that behaves best under the current reconstruction algorithm, which is then taken as input for the inner problem to further optimize the reconstruction method. 
Several heuristic or greedy algorithms \cite{gao2000optimal,seeger2010optimization,chakrabarti2016learning,horstmeyer2017convolutional,gozcu2018learning,cheng2019illumination,haldar2019oedipus,muthumbi2019learned,metzler2020deep} have been proposed to tackle the above nested pair of problems, and has been demonstrated to achieve better reconstructions than conventional methods \cite{muthumbi2019learned,knoll2011adapted} in an empirical study \cite{zijlstra2016evaluation}.
Most recently, in \cite{sherry2020learning}, continuous optimization methods are applied to a formulated bilevel optimization problem, with the aim of learning sparse sampling patterns for MRI within a supervised learning framework for a given variational reconstruction method.
However, the above algorithms suffer from high computational complexity when dealing with large-scale training data, and often lack the flexibility to explore different types of sampling patterns.

\subsection{End-to-end Learning Methods}
\label{sec:comp-with-loupe}
Fueled by the rise of deep learning, some end-to-end data-driven sampling optimization methods \cite{bahadir2019learning,bahadir2020deep,weiss2021pilot,huijben2020learning,aggarwal2020j} are proposed to achieve computational efficiency and improve reconstruction.
PILOT \cite{weiss2021pilot} introduces hardware constraints into the learning pipeline and explores an optimal physically viable $k$-space trajectory that is enabled by interpolation on the discrete $k$-space grid. However, the learned trajectory significantly depends on the initialization and lacks exploration.
J-MoDL \cite{aggarwal2020j} proposes a multichannel sampling model consisting of a non-uniform Fourier operator with continuously defined sampling locations to promote the optimization process without approximations. However, it constrains the sampling mask as the tensor product of two 1D sampling patterns and restricts its performance.
Note that, due to learning a deterministic mask, the above methods inhibit flexibility and randomness, and are against the stochastic strategy of CS theory \cite{candes2006near,duarte2008single}.

LOUPE \cite{bahadir2020deep} assumes that each binary sampling location is an independent Bernoulli random variable and learns a probabilistic sampling pattern instead of a deterministic mask.
In this aspect, both LOUPE and our PUERT have similar inspirations, but in fact, our PUERT enjoys three superior advantages compared to LOUPE. 
First, due to the relaxation of the binarization operation, LOUPE not only pays a performance penalty, but also needs to retrain the reconstruction model with the learned binary mask, whereas our PUERT directly uses the binarization function in the forward pass and introduces an efficient gradient estimation strategy for the backward pass, which overcomes the above two shortcomings and is shown to promote the recovery quality.
Second, LOUPE mainly adopts a U-Net network and does not point out the superiority of DUNs against non-DUNs on further exploitation of the sampling subnet. However, we emphasize that, thanks to DUN's intrinsic structural feature of fully utilizing knowledge on the mask in each stage, it's of great significance to use DUNs to facilitate exploration and training of the sampling subnet.
Third, when testing, LOUPE directly samples from the learned probabilistic pattern and gets a mask with an inexact sampling ratio, whereas in PUERT, the vanilla binarization function is replaced by a greedy version, thus achieving precise control of sampling ratio and promoting fair comparisons.

\begin{figure*}[t]
\centering
\includegraphics[width=0.98\linewidth]{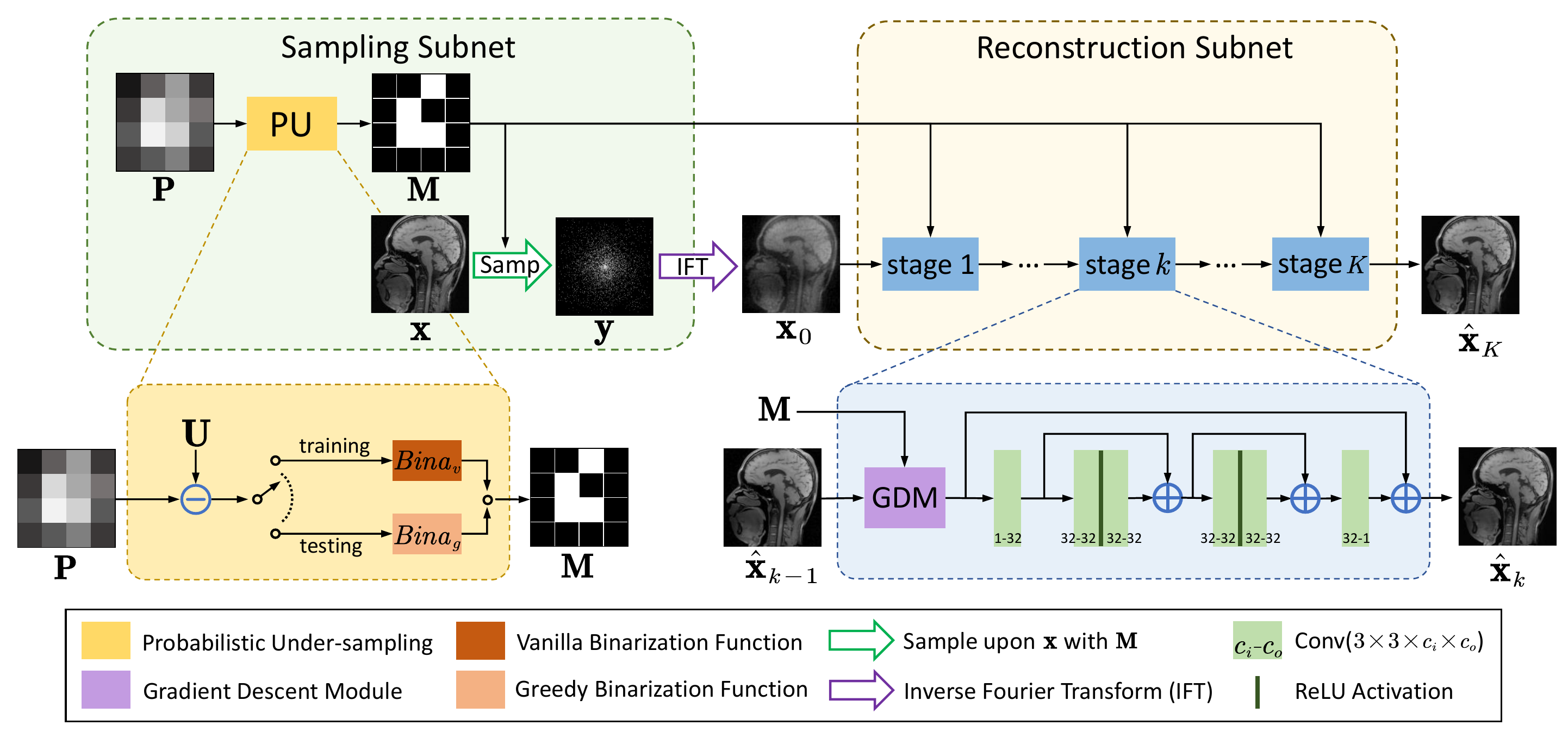}
\vspace{-10pt}
\caption{Our proposed PUERT consists of two subnets: a Sampling Subnet and a Reconstruction Subnet. In the Sampling Subnet, we learn a probabilistic sampling pattern $\mathbf{P}$ to generate the binary mask $\mathbf{M}$.
Each value in $\mathbf{P}$ stands for the probability of the corresponding location in $\mathbf{M}$ of taking value 1.
We generate $\mathbf{M}$ via adopting a uniform-distributed $\mathbf{U}$ and a binarization function to implement Bernoulli sampling. Note that, during testing, we use a greedy binarization function to implement precise control of the sampling ratio.
In the Reconstruction Subnet, we cast the traditional ISTA method into a deep unfolding network containing $K$ stacked stages. Each stage ($k$-th) inputs $\hat{\mathbf{x}}_{k-1}$, outputs $\hat{\mathbf{x}}_{k}$, and is composed of a Gradient Descent Module (GDM) and a Proximal Mapping Module (PMM). We implement PMM with 2 residual blocks, two convolutions and a long residual connection with input.}
\vspace{-6pt}
\label{fig:arch}
\end{figure*} 

\section{Proposed Method}
In this section, we elaborate on the design of our proposed PUERT. We first formulate the problem in Section~\ref{sec:formulation}. Then, as shown in Fig.~\ref{fig:arch}, our PUERT consists of a sampling subnet and a reconstruction subnet, which are described in Section~\ref{sec:sampnet} and Section~\ref{sec:recnet} respectively. Finally, we describe the details about parameters and loss function in Section~\ref{sec:para-loss}.

\subsection{Problem Formulation}
\label{sec:formulation}
In CS-MRI, the acquisition process can be formulated as:
\begin{equation}
\mathbf{y} = \mathbf{M} \odot \mathbf{F} \mathbf{x}+\mathbf{\varepsilon},
\label{eq:acquisition}
\end{equation}
where $\mathbf{x} \in \mathbb{C}^{N_{x} \times N_{y}}$ and $\mathbf{y} \in \mathbb{C}^{N_{x} \times N_{y}}$ denote the original fully-sampled image to be reconstructed and the under-sampled $k$-space observation, respectively.
$\mathbf{\varepsilon} \in \mathbb{C}^{N_{x} \times N_{y}}$ is the noise generated during acquisition.
$\mathbf{F}$ and $\odot$ denote the Fourier Transform (FT) and the element-wise multiplication operation, respectively.
$\mathbf{M} \in \mathbb{R}^{N_{x} \times N_{y}}$ is a \textit{binary} sampling mask matrix composed of 1 and 0, which separately represent whether to sample or not at the corresponding $k$-space position. 
Note that the CS sampling ratio, denoted by $\alpha$, is defined as the ratio of value 1 in $\mathbf{M}$, \textit{i.e.}, $\alpha =\frac{\left \| \mathbf{M} \right \|_{0}}{{N_{x} \times N_{y}}}$.

To tackle the ill-posed inversion of reconstructing $\mathbf{x}$ from sub-sampled data $\mathbf{y}$ under a given mask $\mathbf{M}$, traditional CS-MRI reconstruction methods usually reconstruct the original image $\mathbf{x}$ by solving the following optimization problem:
\begin{equation}
\underset{\mathbf{x}}{\arg \min }\|\mathbf{M} \odot \mathbf{F} \mathbf{x}-\mathbf{y}\|_{2}^{2}+\lambda \psi(\mathbf{x}),
\label{eq:problem}
\end{equation}
where $\psi(\mathbf{x})$ denotes an image prior-regularized term and $\lambda$ is a weight to balance between fidelity and regularization.
Despite the broad task defined above, note that we only consider partial Fourier reconstruction here, excluding parallel imaging \cite{pruessmann1999sense,griswold2002generalized,lustig2010spirit} and structured low-rank matrix methods \cite{shin2014calibrationless,haldar2013low,jin2016general}.

In pursuit of superior recovery performance, we propose to concurrently explore an optimal sampling pattern while learning reconstruction network parameters. To this end, we take $\mathbf{M}$ as learnable parameters and construct a novel end-to-end Probabilistic Under-sampling and Explicable Reconstruction neTwork (PUERT), which is composed of a sampling subnet and a reconstruction subnet, as illustrated in Fig.~\ref{fig:arch}.

\subsection{Sampling Subnet}
\label{sec:sampnet}
In the Sampling Subnet (\textbf{SampNet}), we first explore a Probabilistic Under-sampling (PU) scheme to preserve robustness and stochastics, and then adopt a Dynamic Gradient Estimation (DGE) strategy to enable efficient training and promote network performance.

\vspace{2pt}
\textbf{Probabilistic Under-sampling (PU):}
Building on stochastic strategies and the robustness requirement of compressed sensing, stochastic sub-sampling patterns are able to create noise-like artifacts that are relatively easier to remove \cite{gamper2008compressed}.
Therefore, we propose to directly optimize a probabilistic sampling pattern rather than a fixed binary mask. 
Furthermore, considering that adopting a classic sampling pattern (\textit{e.g.}, Gaussian \cite{wang2009variable} or Uniform \cite{gamper2008compressed} distribution) exhibits great limitation and lacks adaptability, we propose to learn independent Bernoulli random variables for each $k$-space point, thus learning a sampling pattern highly customized to specific training data and recovery methods.

\begin{figure*}[t]
\centering
\includegraphics[width=1\linewidth]{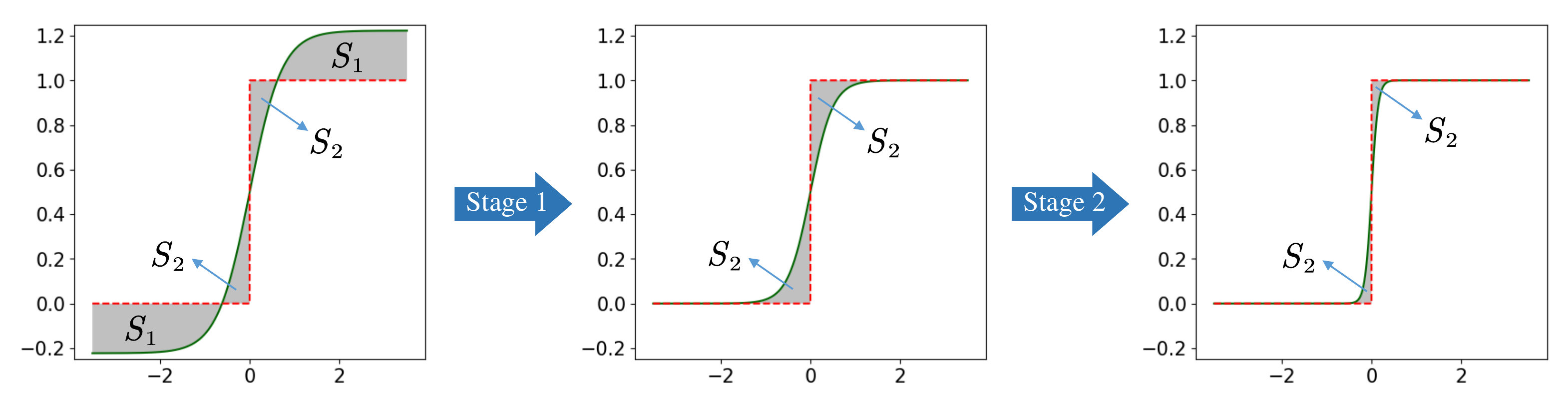}
\vspace{-20pt}
\caption{Our proposed Dynamic Gradient Estimation consists of two stages. The error of gradient estimation is denoted by the area of gray shades. During Stage 1, in order to retain fast updating ability, we keep the derivative value close to one and reduce $S_{1}$ by decreasing the output range. While during Stage 2, more attention is paid to achieving accurate gradients and we increase the slope of the function to make $S_{2}$ shrink.}
\vspace{-4pt}
\label{fig:dge}  
\end{figure*}

Concretely, a learnable probabilistic sampling pattern $\mathbf{P} \in \mathbb{R}^{N_{x} \times N_{y}}$ is introduced and each value $\mathbf{P}_{i,j} \in [0,1]$ stands for the probability of $\mathbf{M}_{i,j}$ taking value 1, \textit{i.e.}, $\mathbf{M}_{i,j} \sim \mathcal{B}(\mathbf{P}_{i,j})$, where $\mathcal{B}(z)$ denotes a Bernoulli random variable with parameter $z$. 
In order to draw a sample from $\mathbf{P}$ and generate a binary mask $\mathbf{M}$, 
we introduce a matrix $\mathbf{U} \in \mathbb{R}^{N_{x} \times N_{y}}$ with each element independently drawn from a Uniform distribution on $[0,1]$, \textit{i.e.}, $\mathbf{U}_{i,j} \sim \mathcal{U}(0,1)$,
and then binarize the difference between $\mathbf{P}_{i,j}$ and $\mathbf{U}_{i,j}$ by a function $Bina$. So the generation of $\mathbf{M}$ is formulated as:
\begin{equation}
\mathbf{M}_{i,j} = Bina(\mathbf{P}_{i,j} - \mathbf{U}_{i,j}), 
\label{eq:Bernoulli-sample}
\end{equation}
where $Bina$ is implemented by a vanilla binarization function $Bina_{v}$ in the \textbf{training} process:
\begin{equation}
Bina_{v}(x)=\left\{\begin{matrix}
 1, & x>=0 \\ 
 0, & x<0.
\end{matrix}\right.
\label{eq:bina-v}
\end{equation}
Note that $\mathbf{U}$ is not fixed, but randomly generated every time $\mathbf{P}$ generates $\mathbf{M}$.
In this way, the sampling subnet is able to learn a probabilistic sampling pattern $\mathbf{P}$ that expresses belief or importance across all $k$-space locations, thus gaining robustness and stochastics for a more reliable CS-MRI reconstruction. 
Note that the use of $\mathbf{U}$ is similar to the re-parameterization trick used in VAE\cite{kingma2013auto}, which generates non-uniform random numbers by transforming some base distribution, so as to recast the statistical expression and tackle the stochastic node.

Another critical issue is how to control the sparsity of the generated mask $\mathbf{M}$ at a given target sampling ratio $\alpha$. To this end, we first introduce a rescale operator to adjust the average value of $\mathbf{P}$, and then adopt a greedy binarization operator specifically for testing to implement a totally accurate control.

To be concrete, before sampling out $\mathbf{M}$ as Eq.~(\ref{eq:Bernoulli-sample}), the probabilistic sampling pattern $\mathbf{P}$ is rescaled as follows:
\begin{equation}
Res(\mathbf{P}_{i,j})=\left\{\begin{array}{l}
\frac{\alpha}{\bar{p}} \mathbf{P}_{i,j}, ~~~~~~~~~~~~~~\text { if } \bar{p} \geq \alpha\\
1-\frac{1-\alpha}{1-\bar{p}}(\mathbf{1}-\mathbf{P}_{i,j}), \text {otherwise}.
\end{array}\right.
\label{eq:rescale-func}
\end{equation}
\vspace{2pt}
Here, $\bar{p}$ stands for the average value of $\mathbf{P}$, \textit{i.e.}, 
$\bar{p} =\frac{\sum_{i,j} {\mathbf{P}_{i,j}}} {{N_{x} \times N_{y}}}$.
It can be proven that Eq.~(\ref{eq:rescale-func}) yields $\mathbf{P}$ with its average value rescaled to the given CS ratio $\alpha$. With this rule, the ratio of the generated binary mask $\mathbf{M}$ would be \textit{close} to the target $\alpha$.

During the testing process, in order to achieve an \textit{accurate} control of the ratio of $\mathbf{M}$, a greedy binarization function $Bina_{g}$ is further utilized to replace the vanilla one $Bina_{v}$ in Eq.~(\ref{eq:Bernoulli-sample}).
Specifically, the greedy binarization function $Bina_{g}$ used during \textbf{testing} is defined as follows:
\begin{equation}
Bina_{g}(x;\mathbf{\Omega},\alpha)=\left\{\begin{matrix}
 1, & ~x>=b(\mathbf{\Omega},\alpha) \\ 
 0, & x<b(\mathbf{\Omega},\alpha),
\end{matrix}\right.
\label{eq:bina-g}
\end{equation}
where $\mathbf{\Omega}$ represents the set $\left \{ \mathbf{P}_{i,j} - \mathbf{U}_{i,j} \right \}$ and 
$b(\mathbf{\Omega},\alpha)$ stands for the $(\alpha \times \left | \mathbf{\Omega} \right |)$-$th$ largest number in the set $\mathbf{\Omega}$. 
In this way, our method precisely controls the ratio of $\mathbf{M}$ and implements a fair comparison with other methods. 
Note that since the learned probabilistic pattern $\mathbf{P}$ represents importance across all $k$-space locations, our value-maximization design as Eq.~(\ref{eq:bina-g}) for the testing process is actually cooperative with the training period and exhibits nice rationality. Besides, thanks to the adoption of $\mathbf{U}$ in Eq.~(\ref{eq:bina-g}), our method outputs a sampling mask with randomness rather than a fixed mask.

\vspace{2pt}
\textbf{Dynamic Gradient Estimation (DGE):}
In order to make the above sampling subnet trainable, inspired by recent works in Binarized Neural Networks (BNNs) \cite{simons2019review,qin2020forward}, a dynamic gradient estimation strategy is further introduced for the vanilla binarization function $Bina_{v}$ in Eq.~(\ref{eq:bina-v}).
To be specific, we adopt the following dynamic function $g$ as a progressive approximation of $Bina_{v}$ during backward propagation:
\begin{equation}
g(x)=\frac{k \tanh ( 2 t x ) +1}{2},
\label{eq:dge}
\end{equation}
whose derivative is used as the backward gradient of $Bina_{v}$.
Here $t$ and $k$ are control variables and change as follows during the training period:
\begin{equation}
t=T_{\min } 10^{\frac{i}{N_{e}} \times \lg \frac{T_{\max }}{T_{\min }}}, \quad k=\max \left(\frac{1}{t}, 1\right),
\end{equation}
where $i$ is the current epoch and $N_{e}$ is the number of epochs, $T_{\min }=10^{-1}$ and $T_{\max }=10^{1}$. Note that $t$ and $k$ controls the slope and output range of function $g$, respectively.

Essentially, this is a progressive two-stage strategy to approximate the vanilla binarization function $Bina_{v}$. 
As shown in Fig.~\ref{fig:dge}, during Stage 1, \textit{i.e.}, the early stage of training, we focus more on the updating ability and speed of the backward propagation algorithm. Therefore, the gradient estimation function’s derivative value is kept close to one, and then the output range is reduced progressively from a large scope to $[0,1]$.  
As for Stage 2, we lay more importance on retaining accurate gradients for parameters around zero. Consequently, we keep the function output range as $[0,1]$ and gradually increase the slope so as to push the estimation curse to the shape of the binarization function $Bina_{v}$, thus achieving the consistency of forward and backward propagation during the late stage of training.
Based on the above two-stage scheme, our proposed DGE gradually approximates the binarization function in backward propagation and reasonably updates all parameters, which is shown to efficiently retain the gradient information and promote network performance.

\begin{figure*}[!t]
\begin{tabular}[b]{ c@{ }  c@{ } c@{ } c@{ }  c@{ } c@{ } c@{ } }
   &  \footnotesize VD-1D & \footnotesize PUERT-1D  & \footnotesize Pseudo Radial & \footnotesize Random Uniform  & \footnotesize VD-2D & \footnotesize  PUERT-2D  \\
    \footnotesize 5\% & 
    \includegraphics[clip,width=.153\textwidth,valign=c]{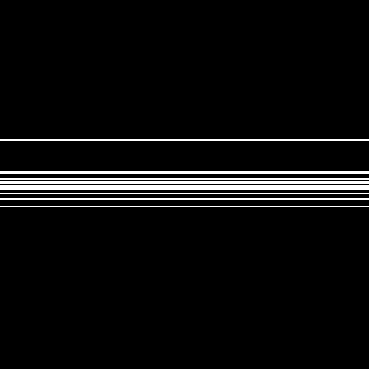}&
  	\includegraphics[clip,width=.153\textwidth,valign=c]{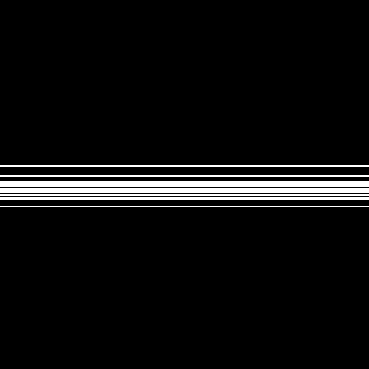}&   
    \includegraphics[clip,width=.153\textwidth,valign=c]{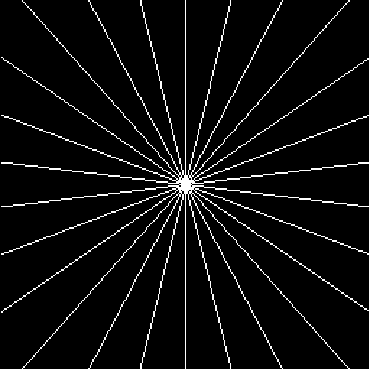}&
    \includegraphics[clip,width=.153\textwidth,valign=c]{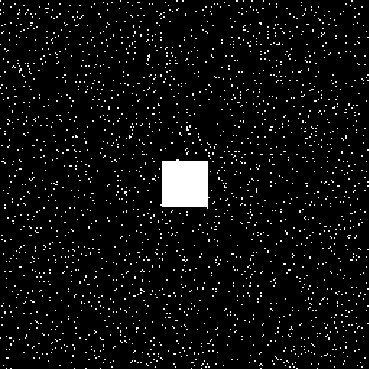}&
    \includegraphics[clip,width=.153\textwidth,valign=c]{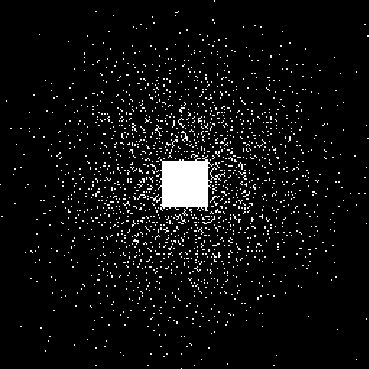}&
    \includegraphics[clip,width=.153\textwidth,valign=c]{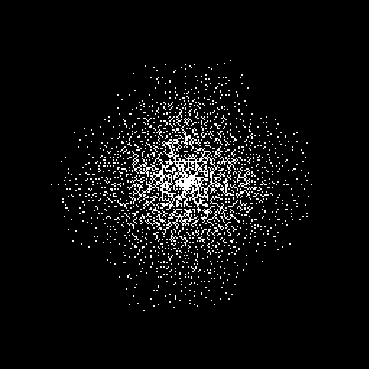}
\\
\\
    \footnotesize 10\% & 
    \includegraphics[clip,width=.153\textwidth,valign=c]{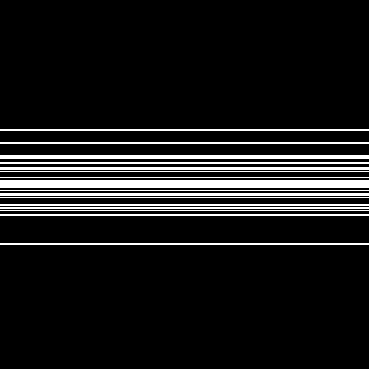}&
  	\includegraphics[clip,width=.153\textwidth,valign=c]{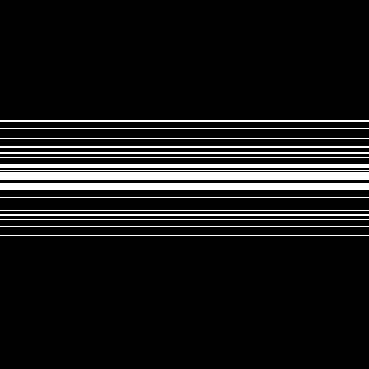}&   
    \includegraphics[clip,width=.153\textwidth,valign=c]{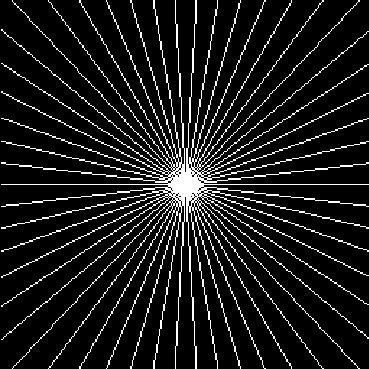}&
    \includegraphics[clip,width=.153\textwidth,valign=c]{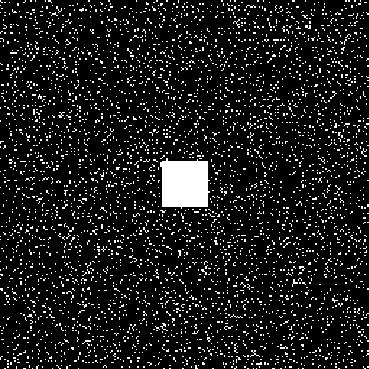}&
    \includegraphics[clip,width=.153\textwidth,valign=c]{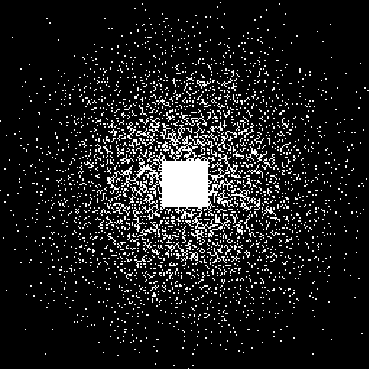}&
    \includegraphics[clip,width=.153\textwidth,valign=c]{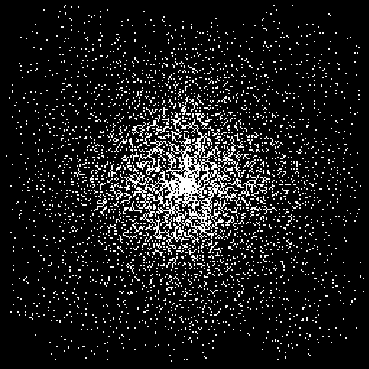}
\\
\end{tabular}
\caption{Visual reconstruction comparisons of different masks under RecNet trained on Brain dataset for two different ratios $\alpha = 5\%$ and $\alpha = 10\%$.
VD is short for Variable Density.
Compared with VD-2D masks, the learned 2D masks are similar for lower frequencies, but different for high frequency values, which demonstrates that our PUERT enjoys the ability to adaptively optimize sub-sampling patterns for specific ratios.}
\vspace{-2pt}
\label{fig:masks}  
\end{figure*}

\subsection{Reconstruction Subnet}
\label{sec:recnet}
In the Reconstruction Subnet (\textbf{RecNet}), without losing generality, we construct a deep unfolding network by casting the traditional ISTA method into a deep network form, so as to achieve simplicity, effectiveness and interpretability.

The classic CS-MRI optimization problem Eq.~(\ref{eq:problem}) can be efficiently solved with ISTA by iterating the following two update steps:
\begin{equation}
\mathbf{r}_{k}=\hat{\mathbf{x}}_{k-1}-\rho \mathbf{F}^{H} \left(\mathbf{M} \odot \mathbf{F} \hat{\mathbf{x}}_{k-1}-\mathbf{y}\right),
\label{eq:ista_rk}
\end{equation}
\begin{equation}
\hat{\mathbf{x}}_{k}=\operatorname{prox}_{\lambda \psi}\left(\mathbf{r}_{k}\right).
\label{eq:ista_xk}
\end{equation}
Here, $k$ denotes the iteration index, $\rho$ is the step size, $\mathbf{F}^{H}$ denotes the Inverse Fourier Transform (IFT) and the proximal mapping operator of regularizer $\psi$ is defined as $\mathtt{prox}_{\lambda\psi}(\mathbf{r})=\arg\min_\mathbf{x}\frac{1}{2}||\mathbf{x}-\mathbf{r}||^2_2+\lambda\psi(\mathbf{x})$. Eq.~(\ref{eq:ista_rk}) and Eq.~(\ref{eq:ista_xk}) are usually called gradient descent step (GDS) and proximal mapping step (PMS), respectively.

Following \cite{zhang2018ista}, our ISTA-unfolding reconstruction subnet consists of $K$ stages and each stage corresponds to one iteration in ISTA. Concretely, each stage is composed of a gradient descent module (GDM) and a proximal mapping module (PMM), which correspond to the above two update steps Eq.~(\ref{eq:ista_rk}) and Eq.~(\ref{eq:ista_xk}), respectively.

For GDM, to preserve the ISTA structure while increasing network flexibility, the step size $\rho$ is allowed to vary across iterations, and GDS, \textit{i.e.}, Eq.~(\ref{eq:ista_rk}) is casted as follows:
\begin{equation}
\label{eq:gdm}
\mathbf{r}_{k}=\hat{\mathbf{x}}_{k-1}-\rho_{k} \mathbf{F}^{H} \left(\mathbf{M} \odot \mathbf{F} \hat{\mathbf{x}}_{k-1}-\mathbf{y}\right).
\end{equation}

As for PMM in each stage, we propose a simple yet effective module as shown in Fig.~\ref{fig:arch}, which can be formulated as:
\begin{equation}
\mathbf{\hat{x}}_{k} =\mathbf{r}_{k}+ \mathcal{H}^{rec}_{k}(\mathcal{H}_{k}^{RB,2}({\mathcal{H}_{k}^{RB,1}}(\mathcal{H}^{ext}_{k}(\mathbf{r}_{k})))).
\label{eq:pmm}
\end{equation}
Here, PMM consists of 2 residual blocks (RBs) $\mathcal{H}_{k}^{RB,1}$ and $\mathcal{H}_{k}^{RB,2}$, two convolution layers $\mathcal{H}^{ext}_{k}$, $\mathcal{H}^{rec}_{k}$, which devote to extracting the image features and reconstruction, respectively, and a long residual connection with input $\mathbf{r}_{k}$.

\subsection{Network Parameters and Loss Function}
\label{sec:para-loss}
Given the training dataset with full-resolution images $\{ \mathbf{x}^{(i)} \}_{i=1}^{N_D}$
, the sampling subnet first uses a learnable sampling mask $\mathbf{M}$ to get simulation measurements $\mathbf{y}^{(i)} = \mathbf{M} \odot \mathbf{F} \mathbf{x}^{(i)}$, and then, with initialization $\mathbf{x}^{(i)}_0 = \mathbf{F}^{H}\mathbf{y}^{(i)}$ as input, the reconstruction subnet outputs the recovery results, aiming to reduce the discrepancy between $\mathbf{x}^{(i)}$ and $\mathbf{\hat{x}}^{(i)}_K$. Therefore, we use the following end-to-end loss function to train our PUERT:
\begin{equation}
\label{eq:loss}
{\mathcal{L}}(\mathbf{\Theta}) =  \frac{1}{{N_DN}}{\sum^{N_D}_{i=1}\| \mathbf{\hat{x}}^{(i)}_K- \mathbf{x}^{(i)}\|_2^2},
\end{equation}
where $N_D$ is the number of training images and $N$ is the size of each image $\mathbf{x}^{(i)}$, \textit{i.e.}, $N = N_{x} \times N_{y}$. $\mathbf{\Theta}$ denotes the learnable parameter set in the two subnets of PUERT.
Note that, in the sampling subnet, the learnable parameters are not directly the probabilistic sampling pattern $\mathbf{P}$, but an unconstrained image $\mathbf{O}$ with the same size, such that $\mathbf{P}_{i,j} = \sigma(\mathbf{O}_{i,j})$ to realize $\mathbf{P}_{i,j} \in (0,1)$. Here, $\sigma$ denotes the sigmoid fuction with the slope set be to 5.
It is also worth emphasizing that, thanks to the rescale operator as Eq.~(\ref{eq:rescale-func}), a loss term to constrain the sparsity ratio of the learned sampling mask is unnecessary.

\begin{figure*}[!t]
\begin{tabular}[b]{ c@{ } }
% \hspace{-4mm}
    \includegraphics[clip,width=0.995\textwidth,valign=c]{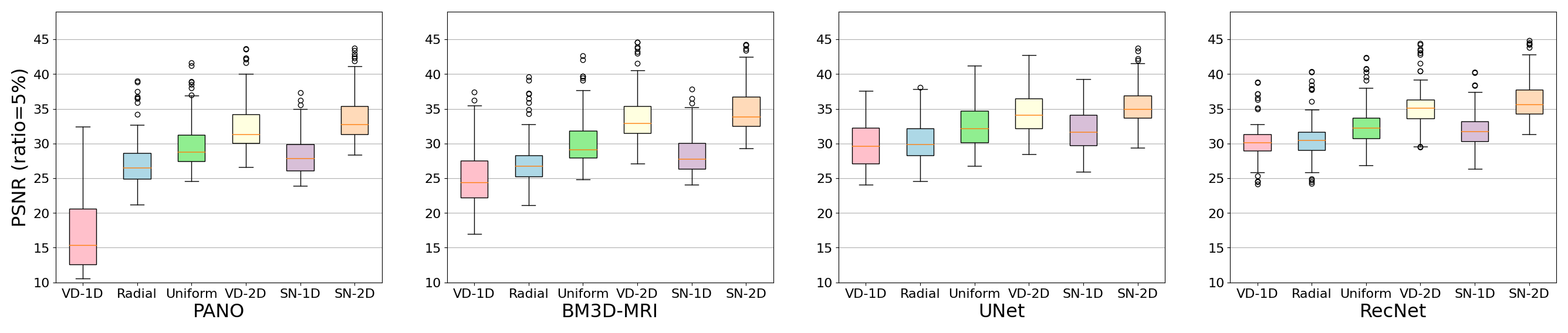} \\
  	\includegraphics[clip,width=0.995\textwidth,valign=c]{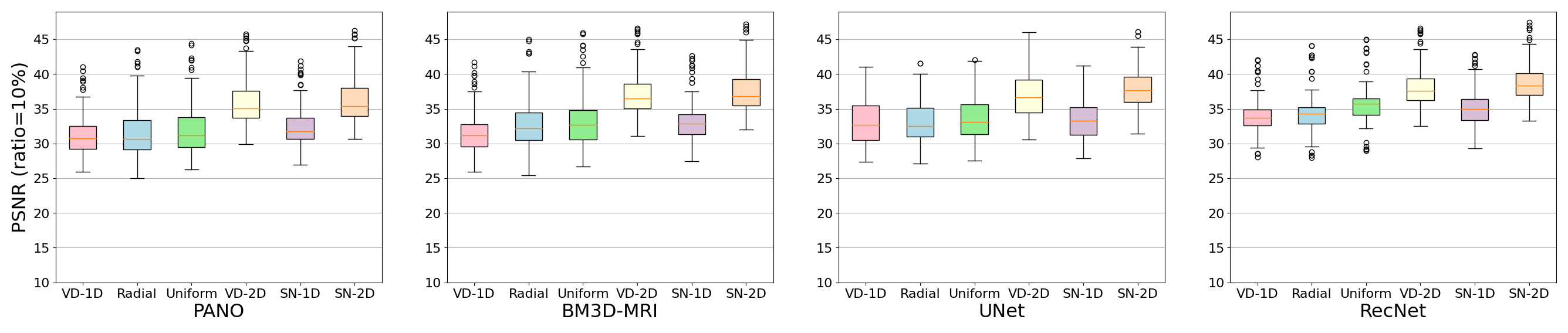} \\
\end{tabular}
\caption{Box plots for PSNR values of reconstructions obtained with four reconstruction methods, six different sampling schemes and two sub-sampling ratios on Brain dataset. SN-1D and SN-2D denote our proposed SampNet under 1D and 2D settings, respectively.
Our SampNet yields attractively better results than competing masks when combined with any of the four reconstruction methods and under any of the two CS ratios.}
\label{fig:box}  
\end{figure*}

\section{Experiments}

Our PUERT is implemented in PyTorch \cite{pytorch} on one NVIDIA Tesla V100. We utilize Adam \cite{DBLP:journals/corr/KingmaB14} optimization with a learning rate of 0.0001 (3000 epochs) and a batch size of 8. The default value of stage number $K$ is 9.
Following common practices in previous works, we use two widely used benchmark datasets: Brain \cite{yang2016deep} and FastMRI \cite{zbontar2018fastmri}, which contain 100 brain and 4501 knee MR images for training, and 50 brain and 657 knee MR images for testing, respectively. 
Brain dataset is based on ground-truth images and follows the simulation process stated in Section~\ref{sec:para-loss}.
As for FastMRI dataset, the simulation is similar, except that the fully-sampled ground truth is in $k$-space (emulated single coil \cite{tygert2020simulating}) instead of the image domain, and thus, is relatively more realistic.
The recovered results are evaluated with Peak Signal-to-Noise Ratio (PSNR) and Structural Similarity Index (SSIM) \cite{wang2004image}.

\subsection{Comparisons with Classic Masks under Multiple Reconstruction Methods}
To validate the effectiveness of our proposed sampling pattern optimization scheme, \textit{i.e.}, our SampNet, we compare our learnable masks with four types of classic fixed masks under the conditions of four reconstruction methods.

\begin{itemize}
\item 
For \textit{classic fixed masks}, we consider four categories that are widely used in the literature: Cartesian \cite{haldar2010compressed} with skipped lines (dubbed VD-1D), pseudo Radial \cite{yang2016deep}, Random Uniform \cite{gamper2008compressed} and Variable Density under 2D (dubbed VD-2D) \cite{wang2009variable} masks, where the first category is 1D sub-sampling mask and the last three are 2D masks. Note that a fixed calibration region of size $32 \times 32$ is adopted in the center of the $k$-space for Uniform and VD-2D masks in order to yield better recovery performance.

\begin{table}[t]
\centering
\footnotesize
\setlength{\tabcolsep}{1.72pt}   
\caption{Average PSNR comparisons of various sampling schemes, with four recovery methods under different CS ratios for both 1D and 2D settings on Brain dataset. Our proposed SampNet yields best results among all competing masks under all conditions.} 
\label{tab:brain-combination}
\begin{tabular}{c|c|cc|cccc}
\shline
\multirow{2}{*}{Ratios} & \multirow{2}{*}{Methods} & \multicolumn{2}{c|}{1D Settings} & \multicolumn{4}{c}{2D Settings}     \\ \cline{3-8} 
                         &                          & VD-1D  & SampNet & Radial  & Uniform & VD-2D  & SampNet \\ \hline
\multirow{4}{*}{5\%}     & PANO \cite{qu2014magnetic} & 17.51            & 28.53         & 27.70  & 30.26   & 33.04    & 34.14  \\ \cline{2-8} 
                        & \scriptsize BM3D-MRI \cite{eksioglu2016decoupled} & 25.43            & 28.72         & 27.78  & 30.77   & 34.26    & 35.27  \\ \cline{2-8} 
                        & U-Net \cite{zbontar2018fastmri} & 30.03            & 32.09         & 30.49  & 32.91   & 34.66    & 35.65  \\ \cline{2-8} 
                        & RecNet                   & 30.64            & 32.26         & 31.02  & 33.00   & 35.79    & 36.74  \\ \hline
\multirow{4}{*}{10\%}    & PANO \cite{qu2014magnetic} & 31.66            & 32.79         & 31.93  & 32.69   & 36.26    & 36.75  \\ \cline{2-8} 
                        & \scriptsize BM3D-MRI \cite{eksioglu2016decoupled} & 32.06            & 33.65         & 33.19  & 33.83   & 37.58    & 38.09  \\ \cline{2-8} 
                        & U-Net \cite{zbontar2018fastmri} & 33.32            & 33.68         & 33.45  & 33.82   & 37.31    & 38.03  \\ \cline{2-8} 
                        & RecNet                  & 34.22            & 35.51         & 34.88  & 36.03   & 38.45    & 39.13  \\ \shline
\end{tabular}
\end{table}

\begin{figure}[!t]
\begin{tabular}[b]{ c@{ }  c@{ } c@{ } c@{ }  }
        &  \footnotesize Example Slices & \footnotesize Optimized under 5\%  & \footnotesize Optimized under 10\% 
\\
    \footnotesize \rotatebox[origin=c]{90}{Knee} & 
    \includegraphics[clip,width=.148\textwidth,valign=c]{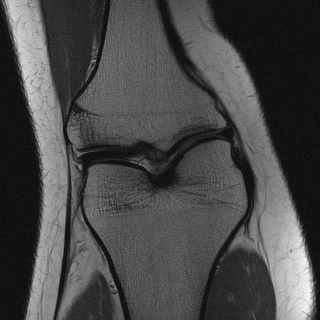}&
  	\includegraphics[clip,width=.148\textwidth,valign=c]{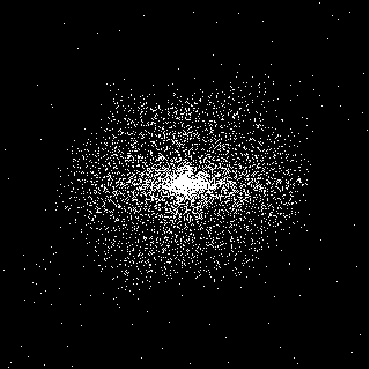}&   
    \includegraphics[clip,width=.148\textwidth,valign=c]{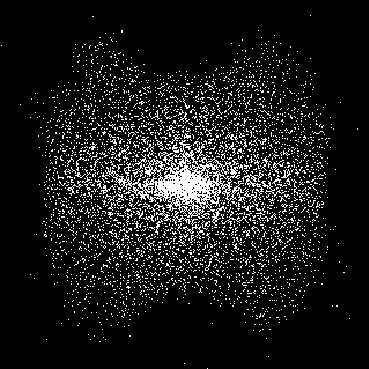}

\\
\\
    \footnotesize \rotatebox[origin=c]{90}{Brain}  & 
    \includegraphics[clip,width=.148\textwidth,valign=c]{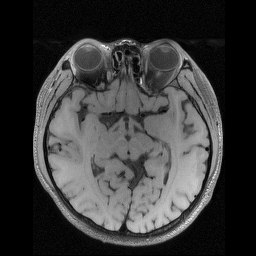}&
  	\includegraphics[clip,width=.148\textwidth,valign=c]{img/masks/mask_cs5_12450_brain.png}&   
    \includegraphics[clip,width=.148\textwidth,valign=c]{img/masks/mask_cs10_8450_brain.png}
\\
\end{tabular}
\vspace{2pt}
\caption{Visual comparisons of PUERT-optimized sampling masks for the knee and brain anatomies (under $\alpha = 5\%$ and $\alpha = 10\%$).
For the knee anatomy, more attention has been paid to lateral frequencies, whereas for the brain, the learned masks are more radially symmetric. This comparison result highlights the importance of adapting the learnable under-sampling pattern to the anatomy and data at hand.}
\label{fig:mask-brain-and-knee}  
\end{figure}

\item 
For \textit{reconstruction methods}, we consider two traditional model-based methods, \textit{i.e.}, PANO \cite{qu2014magnetic} and BM3D-MRI \cite{eksioglu2016decoupled}, and one widely used deep learning model called U-Net \cite{zbontar2018fastmri}. Our reconstruction subnet (RecNet) is also extracted as a recovery model for fair comparisons.

\end{itemize}

\begin{figure*}[!t]
\begin{tabular}[b]{ c@{ }  c@{ } c@{ } c@{ }  c@{ } c@{ }  c@{ } }
   \footnotesize Ground Truth & \footnotesize VD-1D (27.19 dB) & \footnotesize Radial (27.05 dB) & \footnotesize Uniform (28.94 dB) & \footnotesize VD-2D (32.01 dB) & \footnotesize Ours-1D (29.00 dB) & \footnotesize Ours-2D (33.15 dB) \\
 
    \includegraphics[clip,width=.135\textwidth,valign=c]{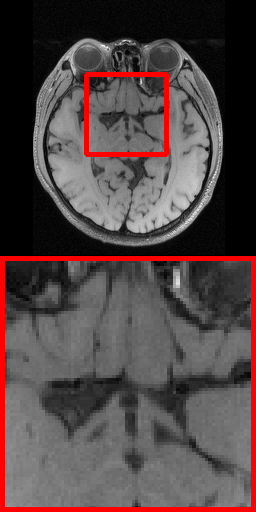}&
  	\includegraphics[clip,width=.135\textwidth,valign=c]{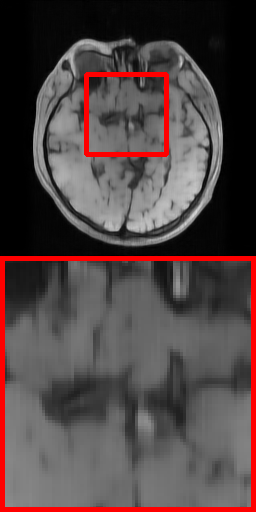}&   
    \includegraphics[clip,width=.135\textwidth,valign=c]{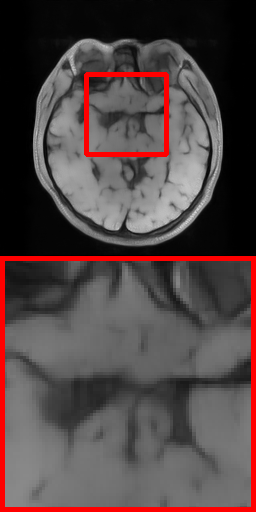}&
    \includegraphics[clip,width=.135\textwidth,valign=c]{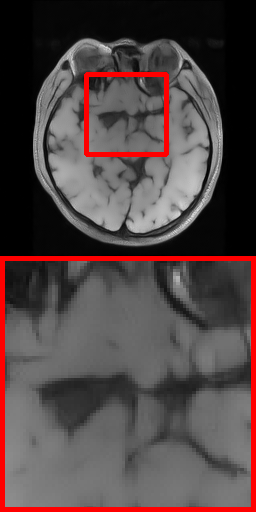}&  
    \includegraphics[clip,width=.135\textwidth,valign=c]{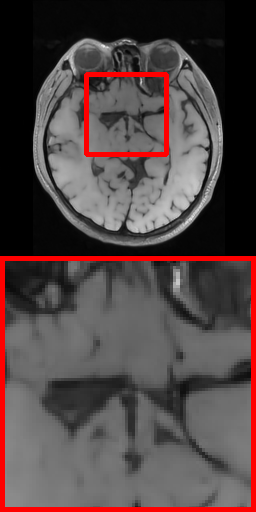}&  
    \includegraphics[clip,width=.135\textwidth,valign=c]{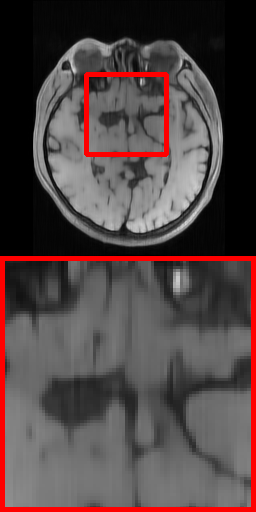}&  
    \includegraphics[clip,width=.135\textwidth,valign=c]{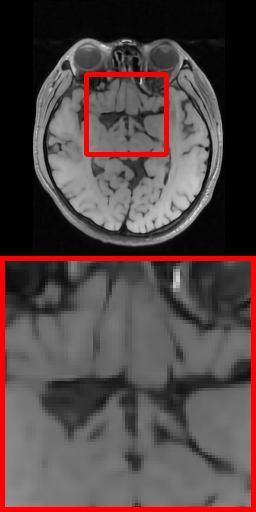}
\\
\\
   \footnotesize Ground Truth & \footnotesize VD-1D (24.62 dB) & \footnotesize Radial (24.24 dB) & \footnotesize Uniform (26.89 dB) & \footnotesize VD-2D (29.47 dB) & \footnotesize Ours-1D (25.97 dB) & \footnotesize Ours-2D (31.72 dB) \\
 
    \includegraphics[clip,width=.135\textwidth,valign=c]{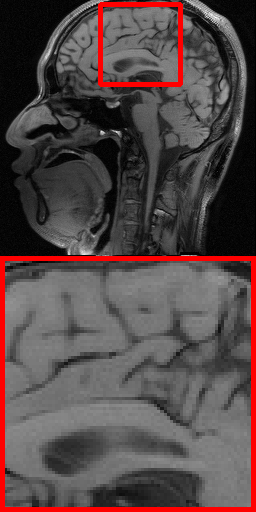}&
  	\includegraphics[clip,width=.135\textwidth,valign=c]{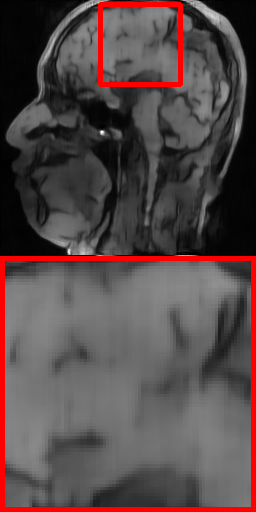}&   
    \includegraphics[clip,width=.135\textwidth,valign=c]{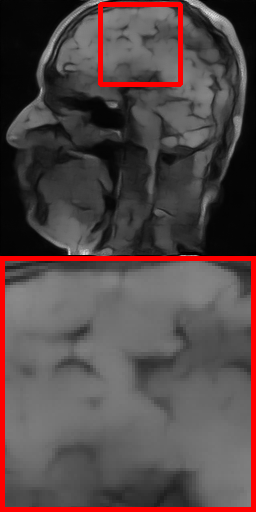}&
    \includegraphics[clip,width=.135\textwidth,valign=c]{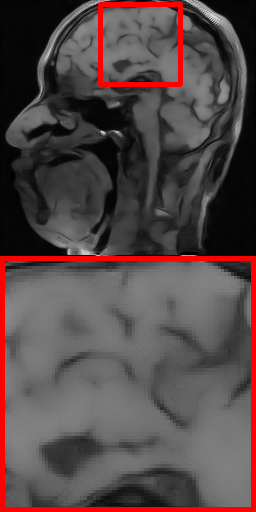}&  
    \includegraphics[clip,width=.135\textwidth,valign=c]{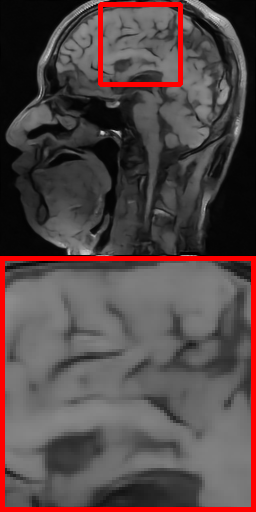}&  
    \includegraphics[clip,width=.135\textwidth,valign=c]{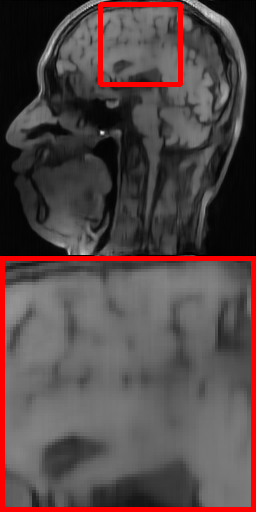}&  
    \includegraphics[clip,width=.135\textwidth,valign=c]{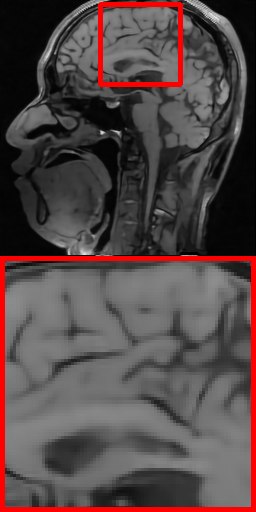}
\\
\end{tabular}
\caption{Visual reconstruction comparisons of different masks under RecNet at ratio 5\% for 1D and 2D settings on Brain dataset. The proposed PUERT is able to recover more details and much sharper edges than other competing masks, even under the aggressive 5\% sampling ratio.}
\label{fig:masks-recimg2}  
\end{figure*}

In this experiment, the aforementioned four classic masks as well as our proposed SampNet,
including 1D and 2D versions, are experimented under the above four reconstruction methods, thus leading to 24 possible combinations.
Note that when combining the two traditional methods PANO and BM3D-MRI with our SampNet, we use the \textit{fixed} sampling masks learned by our PUERT to conduct the reconstruction.
While for U-Net, it is combined with our proposed SampNet to \textit{jointly} optimize the parameters for under-sampling and reconstruction.
Note that for the 1D setting, our SampNet adopts a Bernoulli probabilistic pattern $\mathbf{P} \in \mathbb{R}^{N_{x}}$ and generates a binary vector $\mathbf{V} \in \mathbb{R}^{N_{x}}$, \textit{i.e.}, $\mathbf{V}_{i} \sim \mathcal{B}(\mathbf{P}_{i})$, which is then expanded on the column dimension to a 1D sampling mask $\mathbf{M} \in \mathbb{R}^{N_{x} \times N_{y}}$.

Fig.~\ref{fig:masks} shows the visual comparisons among the classic masks and our optimized sub-sampling matrices trained on Brain dataset for two different ratios $\alpha = 5\%$ and $\alpha = 10\%$. 
One can see that the learned 2D masks are similar to VD-2D masks, both showing a strong preference for lower frequencies and exhibiting a denser sampling pattern closer to the origin of $k$-space. While for high frequency values, compared with VD-2D masks, our optimized masks show a relatively smaller density when $\alpha = 5\%$ but larger when $\alpha = 10\%$, which demonstrates that our PUERT enjoys the ability to adaptively optimize sub-sampling patterns for specific ratios.

Furthermore, the optimized masks by our PUERT for the knee and brain anatomies are compared in Fig.~\ref{fig:mask-brain-and-knee}.
We observed that, for the knee MR images, more attention has been paid to lateral frequencies rather than ventral frequencies, owing to the unique asymmetric feature of the knee anatomy, where there is dramatically more tissue contrast in the lateral direction. The masks learned on Brain dataset, on the other hand, exhibit a more radially symmetric feature. 
This comparison highlights the importance of adapting the learnable under-sampling pattern to the anatomy and data at hand.

Fig.~\ref{fig:box} shows box plots for PSNR values of reconstructions obtained with four reconstruction methods, six different sampling schemes, and two sub-sampling ratios on Brain dataset.
One can intuitively observe that overall our proposed SampNet yields attractively better results than competing masks when combined with any of the four reconstruction methods and under any of the two CS ratios, which demonstrates the superiority of our proposed sub-sampling learning scheme.
It is worth emphasizing that U-Net with our proposed SampNet achieves higher PSNR than competing classic masks, thus validating that the proposed PUERT enjoys the generality of being extended to other reconstruction networks.

Table \ref{tab:brain-combination} lists the concrete PSNR values of Fig.~\ref{fig:box}. 
We can observe that, benefiting from our efficient probabilistic under-sampling scheme and dynamic gradient estimation strategy, the proposed SampNet achieves the highest PSNR results among all competing masks.
Taking RecNet under ratio 10\% for 2D setting as an example, our SampNet achieves remarkable 0.68 dB PSNR gains over the state-of-the-art classic VD-2D masks.
It is also noteworthy that, when equipped with our SampNet, U-Net still performs worse than our RecNet, thus demonstrating that, compared with the classic data-driven reconstruction model, a deep unfolding network is more suitable to facilitate exploration and efficient training of our SampNet.
We attribute such superiority to its intrinsic structural feature of fully utilizing knowledge on the sampling mask in each stage of the deep unfolding network.

Fig.~\ref{fig:masks-recimg2} further shows the visual reconstruction comparisons of different masks under RecNet at ratio 5\% for 1D and 2D settings on Brain dataset. 
As can be intuitively appreciated from the two test images, the PUERT is able to consistently recover more details and much sharper edges than other competing masks, even under the aggressive 5\% sampling ratio, thus validating the effectiveness, efficiency and practicability of our proposed PUERT.

\begin{table*}[t]
\centering
\footnotesize
\setlength{\tabcolsep}{9.6pt} 
\caption{Average PSNR performance comparisons with various state-of-the-art methods with different CS ratios on Brain and FastMRI datasets. The best and second best results are highlighted in \textcolor[rgb]{1.00,0.00,0.00}{red} and \textcolor[rgb]{0.00,0.00,1.00}{blue} colors, respectively.}
\label{tab:sota}
\begin{tabular}{c|c|c|c|c|c|c|c}
\shline
\multirow{2}{*}{Datasets} & \multirow{2}{*}{Methods} & \multirow{2}{*}{Masks}   & \multicolumn{3}{c|}{Ratios} & \\ \cline{4-6} 
                          &                          &                          & 5\%     & 10\%    & 15\%  & \multirow{-2}{*}{Avg.} & \multicolumn{1}{c}{\multirow{-2}{*}{GPU Time}}  \\ \hline \hline
\multirow{10}{*}{Brain}    & Zero-filled  & \multirow{5}{*}{\begin{tabular}[c]{@{}c@{}}Fixed \\ (Radial) \end{tabular}}  & 24.22/0.5140 & 26.81/0.6030 & 28.80/0.6713 & 26.61/0.5961 & 0.0038s\\
                          & UNet-DC \cite{hyun2018deep} &                          & 30.07/0.7540 & 32.83/0.8221 & 34.75/0.8639 & 32.55/0.8133 & 0.0199s \\
                          & ADMMNet \cite{yang2016deep} &                          & 30.13/0.7958 & 34.46/0.8972 & 36.83/0.9306 & 33.81/0.8745 & 0.0192s \\
                          & ISTA-Net+ \cite{zhang2018ista} &                       & 30.64/0.8176 & 34.73/0.9052 & 37.07/0.9343 & 34.15/0.8857 & 0.0277s \\ 
                          & RDN \cite{sun2018compressed} &                         & 31.04/0.8056 & 34.62/0.8887 & 36.85/0.9221 & 34.17/0.8721 & 0.0310s \\
                          & CDDN \cite{zheng2019cascaded} &                    & 31.01/0.8163 & 35.21/0.9074 & 37.35/0.9353 & 34.52/0.8863 & 0.0483s \\ \cline{2-8} 
                          & LOUPE-1D \cite{bahadir2020deep} & \multirow{4}{*}{Learned} &  32.11/0.8647 & 35.16/0.9311 & 36.51/0.9447 & 34.59/0.9135 & 0.0536s \\
                          & LOUPE-2D \cite{bahadir2020deep} &                 & \color{blue}35.34/0.9245 & \color{blue}36.88/0.9381 &  \color{blue}38.30/0.9473 & \color{blue}36.84/0.9366 & 0.0699s \\
                          & PUERT-1D (Ours)                &                        & 32.26/0.8839 & 35.51/0.9285 & 37.44/0.9484 & 35.07/0.9203 & 0.0214s \\
                          & PUERT-2D (Ours)          &                              & \color{red}36.74/0.9409 & \color{red}39.13/0.9551 & \color{red}40.84/0.9636 & \color{red}38.90/0.9532 & 0.0578s \\ \hline \hline
\multirow{10}{*}{FastMRI}  & Zero-filled & \multirow{5}{*}{\begin{tabular}[c]{@{}c@{}}Fixed \\ (Catesian) \end{tabular}} & 25.29/0.5759 & 26.31/0.6205 & 27.50/0.6565 & 26.37/0.6176  & 0.0045s \\
                          & UNet-DC \cite{hyun2018deep} &                           & 27.25/0.6347 & 29.19/0.7048 & 30.60/0.7458 & 29.01/0.6951 & 0.0226s \\
                          & ADMMNet \cite{yang2016deep} &                           & 27.32/0.6372 & 29.05/0.6995 & 30.85/0.7506 & 29.07/0.6958 & 0.0227s \\
                          & ISTA-Net+ \cite{zhang2018ista} &                        & 27.50/0.6436 & 29.82/0.7210 & 31.57/0.7657 & 29.63/0.7101 & 0.0241s \\
                          & RDN \cite{sun2018compressed} &                          & 27.70/0.6487 & 30.19/0.7290 & 31.78/0.7721 & 29.89/0.7166 & 0.0252s \\
                          & CDDN \cite{zheng2019cascaded} &                     & 27.81/0.6507 & 30.32/0.7272 & 31.92/0.7715 & 30.02/0.7165 & 0.0723s \\ \cline{2-8}
                          & LOUPE-1D \cite{bahadir2020deep} &  \multirow{4}{*}{Learned} & 30.68/0.7217 & 31.15/0.7689 & 33.06/0.8211 & 31.63/0.7706 & 0.0487s \\
                          & LOUPE-2D \cite{bahadir2020deep} &                     & \color{blue}31.94/0.7530 & {\color{blue}33.43}/\color{red}0.8033 & {\color{blue}34.73}/\color{red}0.8409 & {\color{blue}33.37}/\color{red}0.7991 & 0.0623s \\
                          & PUERT-1D (Ours)                &                          & 30.71/0.7170 & 32.57/0.7712 & 33.75/0.8110 & 32.34/0.7664 & 0.0301s \\
                          & PUERT-2D (Ours)                &                          & \color{red}32.87/0.7557 & {\color{red}33.96}/\color{blue}0.7991 &  {\color{red}35.16}/\color{blue}0.8412 & {\color{red}34.00}/\color{blue}0.7987 & 0.0694s \\ \shline
\end{tabular}
\end{table*}

\subsection{Comparisons with State-of-the-Art Methods}

We compare our proposed PUERT (including 1D and 2D versions) with six representative state-of-the-art methods, namely UNet-DC \cite{hyun2018deep}, ADMMNet \cite{yang2016deep}, ISTA-Net \cite{zhang2018ista}, RDN \cite{sun2018compressed}, CDDN \cite{zheng2019cascaded} and LOUPE \cite{bahadir2020deep}. The first five methods are reconstruction networks trained under some \textit{fixed} sampling mask, while the last method jointly optimizes the \textit{learnable} sampling pattern and the reconstruction network parameters. Following \cite{yang2016deep} and \cite{zbontar2018fastmri}, for those trained under some fixed mask, we adopt Radial masks (2D) and Cartesian masks with skipped lines (1D) on Brain and FastMRI datasets respectively. 
Note that LOUPE is also able to implement both 1D and 2D sub-sampling optimization schemes.

The average PSNR/SSIM performance reconstructions of various methods on two datasets with respect to three CS ratios are summarized in Table ~\ref{tab:sota}. 
One can observe that methods with learnable sampling patterns, especially our PUERT, overall yield higher PSNR than those with fixed masks, which verifies the superiority of sampling pattern optimization schemes.
Concretely, for Brain dataset, our PUERT-1D and PUERT-2D achieve on average 0.55 dB and 4.38 dB PSNR gain over the state-of-the-art method CDDN respectively. It is noteworthy that the Radial masks used for Brain dataset belong to a 2D setting, whereas our PUERT under the 1D setting still enjoys superiority by a large margin. As for FastMRI dataset, our PUERT-1D and PUERT-2D achieve on average 2.32 dB and 3.98 dB PSNR gain over the state-of-the-art method CDDN respectively.

When compared with the sampling pattern optimization scheme LOUPE, one can observe that, on FastMRI dataset, LOUPE-2D has a slightly better SSIM while PUERT-2D obtains a remarkably higher PSNR on average.
As for Brain dataset, the proposed PUERT-2D consistently produces higher PSNR and SSIM results than LOUPE-2D across three CS ratios. 
The performance advantage of PUERT under 2D is also shown under 1D.
The above results verify the superiority of the proposed PUERT against LOUPE.
Note that, LOUPE mainly adopts the U-Net architecture as the recovery network, it also shows an example of reconstructing with a DUN called CascadeNet \cite{schlemper2017deep, bahadir2020deep}, with an observation that the performance metrics are very close between different reconstruction networks. 
However, we highlight the importance of adopting DUNs to facilitate exploration and efficient training of the sampling subnet, owing to the intrinsic structural feature of fully utilizing knowledge on the sampling mask.
More discussions on the importance of DUN are shown in Section~\ref{sec:dun}.

As shown in Table~\ref{tab:sota}, six representative state-of-the-art methods are based on neural networks, and thus achieve a real-time reconstruction speed, with the inference GPU time less then 0.1s. Among the five methods trained under some fixed mask, CDDN needs the largest amount of inference time, owing to the dense blocks and dilated convolutions adopted in the network. Besides, compared with LOUPE, our PUERT achieves a comparable inference speed on average, but with higher PSNR in reconstruction accuracy, thus demonstrating the efficiency and superiority of the proposed PUERT.

\begin{figure*}[!t]
\begin{tabular}[b]{ c@{ }  c@{ } c@{ } c@{ }  c@{ } c@{ }  c@{ } }
   \footnotesize Ground Truth & \footnotesize ISTA-Net (29.77 dB) & \footnotesize RDN (29.97 dB) & \footnotesize CDDN (30.10 dB) & \footnotesize LOUPE (32.48 dB) & \footnotesize Ours-1D (32.21 dB) & \footnotesize Ours-2D (32.95 dB) \\
 
    \includegraphics[clip,width=.135\textwidth,valign=c]{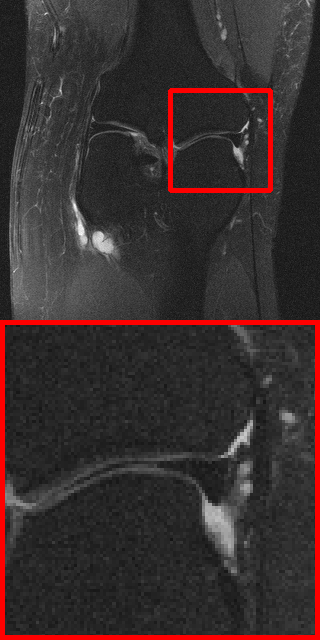}&
  	\includegraphics[clip,width=.135\textwidth,valign=c]{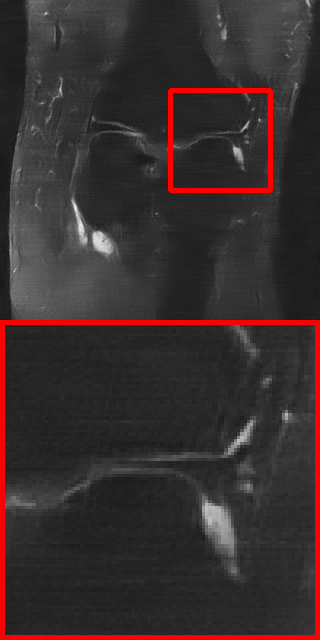}&   
    \includegraphics[clip,width=.135\textwidth,valign=c]{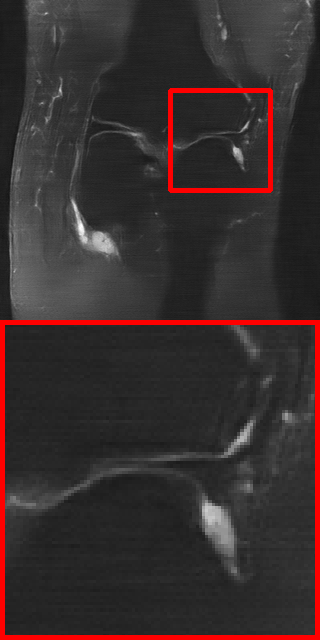}&
    \includegraphics[clip,width=.135\textwidth,valign=c]{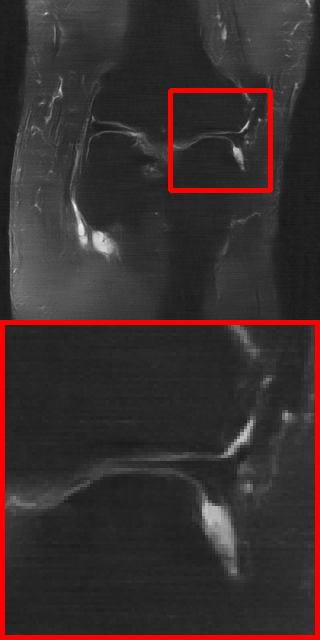}&  
    \includegraphics[clip,width=.135\textwidth,valign=c]{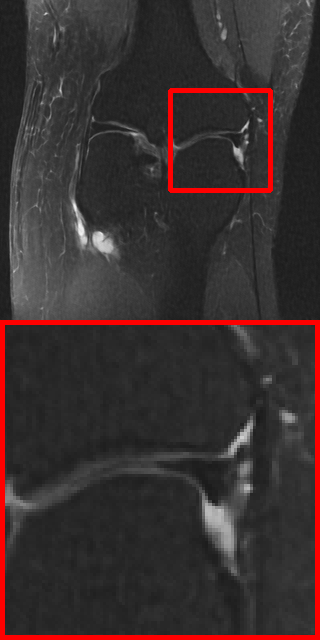}&  
    \includegraphics[clip,width=.135\textwidth,valign=c]{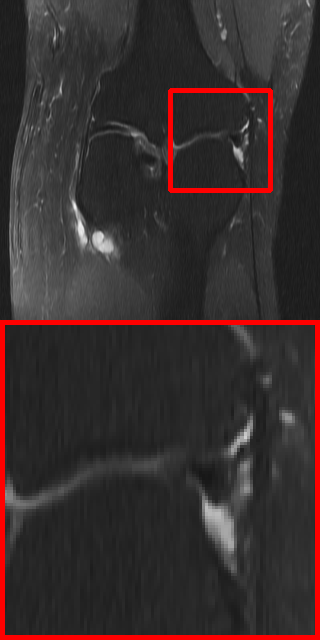}&  
    \includegraphics[clip,width=.135\textwidth,valign=c]{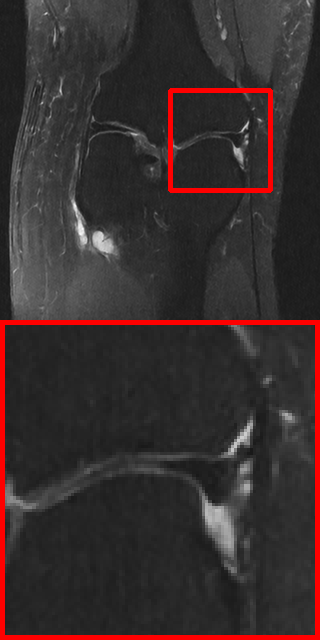}
\\
\\
    \footnotesize Ground Truth & \footnotesize ISTA-Net (29.57 dB) & \footnotesize RDN (29.64 dB) & \footnotesize CDDN (29.91 dB) & \footnotesize LOUPE (33.57 dB) & \footnotesize Ours-1D (32.53 dB) & \footnotesize Ours-2D (35.15 dB) \\
 
    \includegraphics[clip,width=.135\textwidth,valign=c]{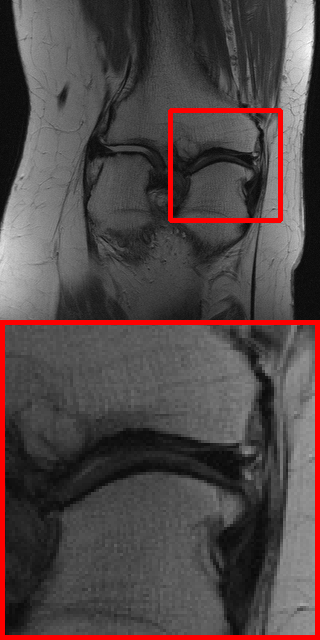}&
  	\includegraphics[clip,width=.135\textwidth,valign=c]{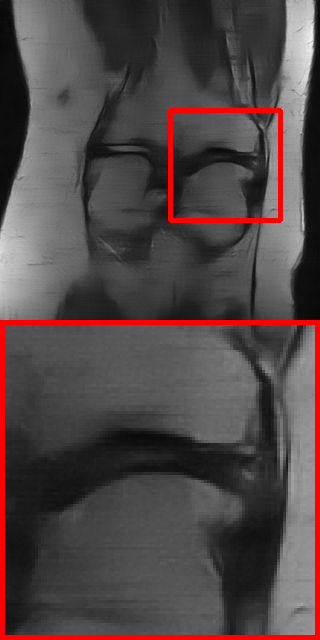}&   
    \includegraphics[clip,width=.135\textwidth,valign=c]{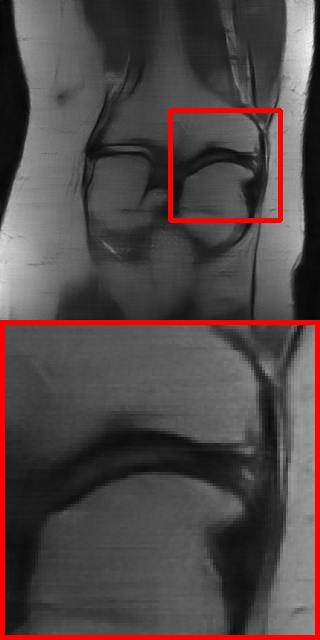}&
    \includegraphics[clip,width=.135\textwidth,valign=c]{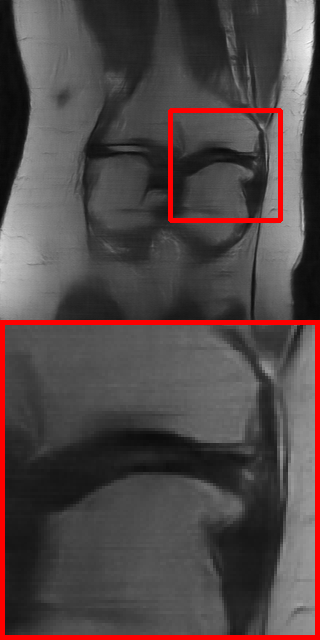}&  
    \includegraphics[clip,width=.135\textwidth,valign=c]{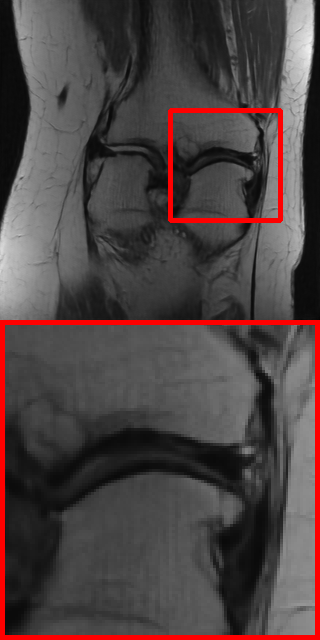}&  
    \includegraphics[clip,width=.135\textwidth,valign=c]{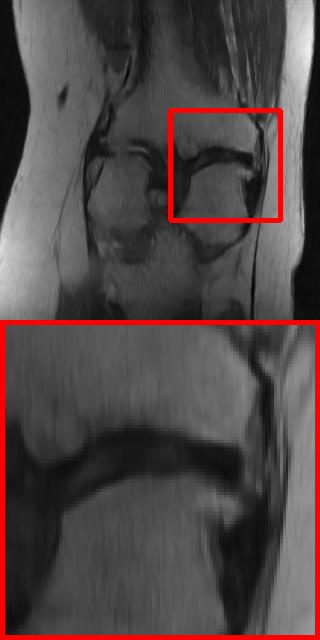}&  
    \includegraphics[clip,width=.135\textwidth,valign=c]{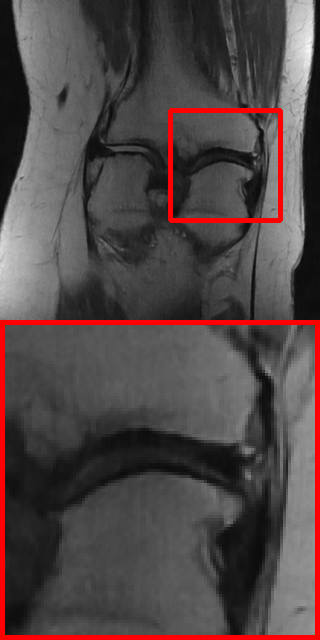}
\\
\end{tabular}
\caption{Visual reconstruction comparisons with various state-of-the-art methods on FastMRI dataset with CS ratio $\alpha = 10\%$. Obviously, the proposed PUERT is able to produce more faithful and clearer results than the other competitive methods.}
\label{fig:rec-knee}  
\end{figure*}

Furthermore, visual comparisons of all the competing methods on FastMRI dataset are shown in Fig.~\ref{fig:rec-knee} with CS ratio $\alpha = 10\%$. Obviously, the proposed PUERT is able to produce more faithful and clearer results than the other competitive methods. 
Overall, the above experiments on two widely used MRI datasets demonstrate that our proposed PUERT performs favorably against state-of-the-arts in terms of both quantitative metrics and visual quality.

\begin{table}[t]
\centering
\footnotesize
\setlength{\tabcolsep}{5.5pt}   
\caption{Average PSNR results for ablation study on PU when ratio is 5\% and 10\% under 1D and 2D settings on Brain dataset.}
\label{tab:pu}
\begin{tabular}{c|c|c|c|c|c|c}
\shline
\multirow{2}{*}{Methods} & \multirow{2}{*}{$\mathbf{PU}$} & \multirow{2}{*}{$\mathbf{U}$ (test)} & \multicolumn{2}{c|}{1D setting} & \multicolumn{2}{c}{2D setting} \\ \cline{4-7} 
       &   &  & 5\%    & 10\%       & 5\%      & 10\%   \\ \hline \hline
case (a) & \ding{55}  & \ding{55} & 31.02  & 35.06  & 36.60  & 39.09 \\
case (b) & \ding{55}  & \ding{55} & 31.08  & 34.91  & 36.51  & 38.91 \\
case (c) & \ding{51}  & \ding{55} & 32.18  & 35.49  & 36.67  & 39.14 \\
PUERT    & \ding{51}  & \ding{51} & 32.26  & 35.51  & 36.74  & 39.13 \\ \shline
\end{tabular}
\end{table}

\subsection{Ablation on Probabilistic Under-Sampling}
In this section, we present an ablation study on Probabilistic Under-Sampling (PU) in our sampling subnet, so as to emphasize the importance of learning a probabilistic sampling pattern rather than a deterministic mask.

\begin{figure}[t]
\centering
\includegraphics[width=0.84\linewidth]{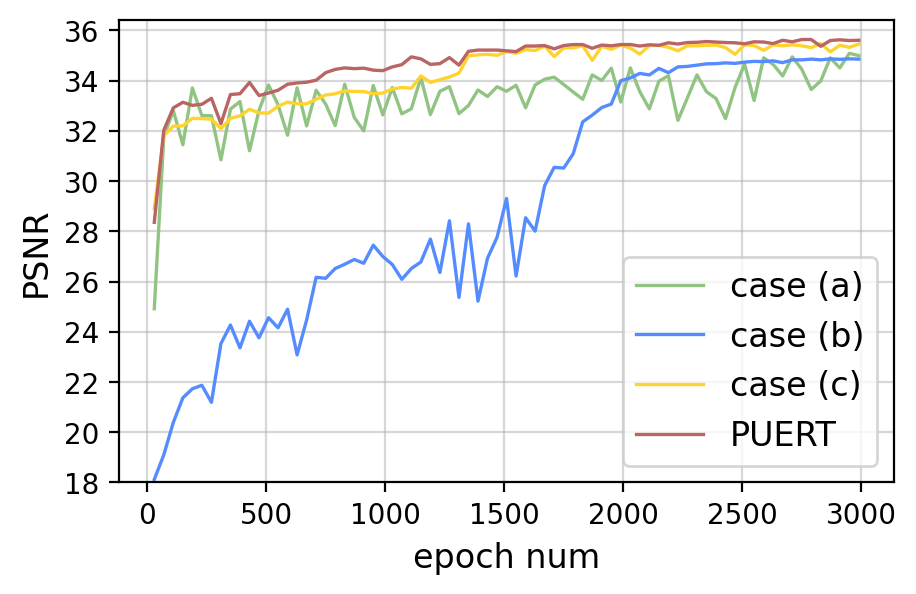}
\vspace{-8pt}
\caption{The progression curves of test PSNR results for ablation study on PU when ratio is 10\% under the 1D setting on Brain dataset.}
\label{fig:pu}  
\end{figure}

First, we consider two possible mask learning schemes uncorrelated with probability (without PU), named case (a) and (b), so as to demonstrate the superiority of PU during \textit{training}.
Then, we consider testing without the Uniform distribution $\mathbf{U}$, called case (c), to investigate the contribution of adopting $\mathbf{U}$ during \textit{testing}.
Table~\ref{tab:pu} shows the PSNR comparisons for the above three cases and our PUERT when ratio is 5\% and 10\% under 1D and 2D settings on Brain dataset.
Fig.~\ref{fig:pu} further illustrates their progression curves of test PSNR when ratio is 10\% under the 1D setting.
We detail the above three cases and their results in the following three paragraphs.

In case (a), the sampling subnet directly binarizes the unconstrained parameter matrix $\mathbf{O}$ into the sampling mask $\mathbf{M}$ via $\mathbf{M}_{i,j} = Bina(\mathbf{O}_{i,j})$ as a replacement of Eq.~(\ref{eq:Bernoulli-sample}). 
Here, a L2 loss term to constrain the sparsity ratio of the learned sampling mask is necessary. However, such loss term still can not stabilize the ratio to the target setup, and therefore, in pursuit of fair comparisons, we have to test case (a) via greedy binarization Eq.~(\ref{eq:bina-g}) but with $\mathbf{\Omega}$ being the set $\{ \mathbf{O}_{i,j}\}$.
Table~\ref{tab:pu} shows that the final PSNR result of case (a) is inferior against PUERT under all conditions.
This is mainly due to: 1) the limited optimization space without PU, 2) the mismatch between training and testing, and 3) the fluctuation caused by the poor ratio control, as shown in Fig.~\ref{fig:pu}.

In order to eliminate the influence of the latter two limitations in case (a), we design case (b) to adopt greedy binarization during both training and testing, via Eq.~(\ref{eq:bina-g}) with $\mathbf{\Omega}$ being the set $\{ \mathbf{O}_{i,j}\}$.
Table~\ref{tab:pu} reports that case (b) achieves PSNR scores comparable with case (a), but still lower than PUERT.
Fig.~\ref{fig:pu} further shows that, during the later period of training, the curve stability of case (b) is better than case (a). However, in the earlier training period, the PNSR upgrade of case (b) is obviously slow.
We attribute such slow update speed to the greedy binarization used for training, which adopts a movable and relative boundary, rather than a fixed boundary adopted in vanilla binarization (\textit{i.e.}, the constant 1).

With the aim of investigating the contribution of adopting $\mathbf{U}$ during \textit{testing}, we consider case (c) which tests PUERT without the Uniform distribution $\mathbf{U}$, \textit{i.e.}, testing via Eq.~(\ref{eq:bina-g}) but with $\mathbf{\Omega}$ being the set $\{ \mathbf{P}_{i,j}\}$.
Without the randomness and exploration introduced by $\mathbf{U}$, case (c) completely depends on the values in $\mathbf{P}$, and thus, with an earlier $\mathbf{P}$ which has not yet fully learned the probability value, case (c) achieves obviously lower PSNR than PUERT in the earlier period, as shown in Fig.~\ref{fig:pu}.
Table~\ref{tab:pu} also reports that case (c) achieves slightly lower PSNR than PUERT.
Note that, in actual scanning, we might use a fixed mask as case (c), but this experiment still verifies the contribution of adopting randomness.

Overall, the above results corroborate the importance of learning a probabilistic sampling pattern rather than a deterministic mask, thus demonstrating the superiority of our design on Probabilistic Under-Sampling (PU) in the sampling subnet.

\subsection{Ablation on Dynamic Gradient Estimation}

\begin{table}[t]
\centering
\footnotesize
\setlength{\tabcolsep}{5.5pt}   
\caption{Average PSNR results for ablation study on DGE when ratio is 5\% and 10\% under 1D and 2D settings on Brain dataset.}
\label{tab:dge}
\begin{tabular}{c|c|c|c|c|c|c}
\shline
\multirow{2}{*}{Methods} & \multirow{2}{*}{\makecell[c]{Update \\ Speed}} & \multirow{2}{*}{\makecell[c]{Estimation \\ Accuracy}} & \multicolumn{2}{c|}{1D setting} & \multicolumn{2}{c}{2D setting} \\ \cline{4-7} 
       &   &  & 5\%    & 10\%       & 5\%      & 10\%   \\ \hline \hline
STE & \ding{51}  & \ding{55} & 30.22 & 34.38  & 36.25 & 38.06 \\
Sigmoid & \ding{55}  & \ding{51} & 31.08 & 34.93 & 36.64 & 39.04 \\
DGE    & \ding{51}  & \ding{51} & 32.26  & 35.51  & 36.74  & 39.13 \\ \shline
\end{tabular}
\end{table}

In this section, we present an ablation study on our proposed Dynamic Gradient Estimation (DGE) in our sampling subnet, so as to underscore the great significance of two-step gradient estimation.
Note that, in LOUPE, the binarization operator is directly relaxed to the sigmoid function to enable optimizing the network by back-propagation. 
However, this relaxation not only causes a network performance penalty, but also leads to a non-binary output mask, which means that, after one has obtained the optimized non-binary mask, the reconstruction network needs to be retrained with the learned binary mask.
To overcome the above two limitations, our PUERT chooses to still adopt the binarization in the forward pass, but develops an efficient gradient estimator for the backward pass.

Such gradient estimators have been widely investigated in Binarized Neural Networks (BNNs) \cite{simons2019review, yin2019understanding}.
One typical and simple estimator is Straight-Through Estimator (STE), which was first proposed by Hinton \cite{hinton2012neural} to train networks with binary activations (\textit{i.e.}, binary neuron). 
In STE, the values are passed through a binarization layer that evaluates the sign in the forward pass and performs the identity function during the backward pass.
Adopting STE as an ablation study, we replace the dynamic gradient estimation function $g(x)$ in Eq.~(\ref{eq:dge}) simply by the identity function $h(x) = x$ and conduct an experiment.
Besides, we also consider an STE variant proposed in \cite{bengio2013estimating}, which uses the sigmoid function $\sigma (x) =  \frac {1}{1+e^{-x}} $ as a replacement of the identity function.

\begin{figure}[t]
\centering
\includegraphics[width=0.84\linewidth]{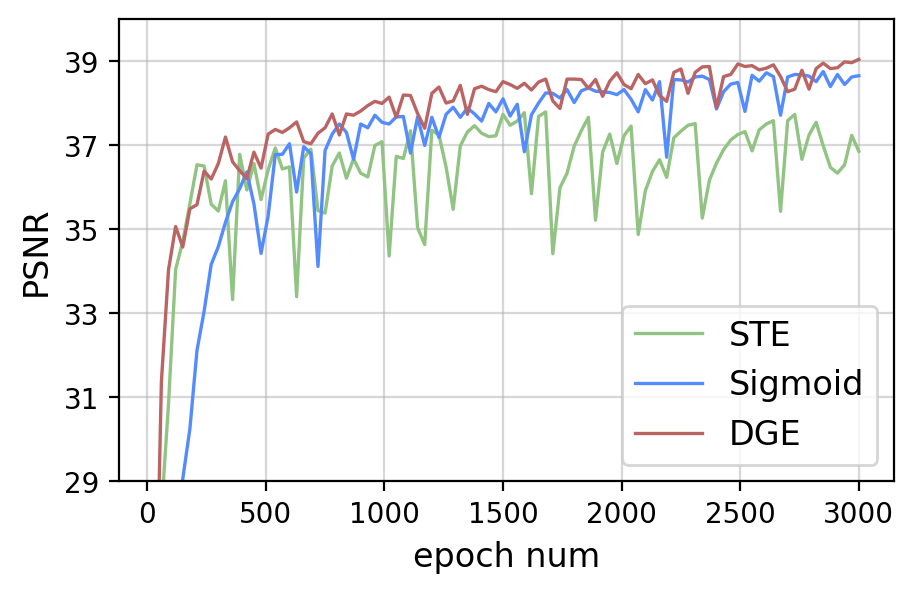}
\vspace{-8pt}
\caption{The progression curves of test PSNR results for ablation study on DGE when ratio is 10\% under the 2D setting on Brain dataset.}
\label{fig:back}  
\end{figure}

Table~\ref{tab:pu} reports the PSNR comparisons for 1) STE, 2) Sigmoid, and 3) our DGE, when ratio is 5\% and 10\% under 1D and 2D settings on Brain dataset, which shows that DGE outperforms the other two methods across all situations. 
Fig.~\ref{fig:back} further provides the progression curves of test PSNR results for the above three methods, when ratio is 10\% under the 2D setting.
One can intuitively observe that, STE converges faster than Sigmoid during the earlier training process, whereas lacks the ability to stably and consistently improve the reconstruction accuracy (resembling Sigmoid) during the later training period.
Such different performances are the result of their different focuses.
Concretely, STE directly passes the gradients through the binarization operator so as to guarantee the updating speed, while Sigmoid estimates the gradients with high similarity to the binarization operator to ensure estimation accuracy.
As a combination of the above two focuses, our proposed two-step DGE is able to dynamically estimate the gradients and emphasize different priorities at the right time, thus implementing both fast convergence speed and high reconstruction accuracy, as can be appreciated from Fig.~\ref{fig:back}.

\subsection{Discussions on the Importance of DUN}
\label{sec:dun}
In the previous statements, we have emphasized to adopt Deep Unfolding Networks (DUNs) for our PUERT in order to fully utilize the knowledge on the learned sampling mask.
In this section, two experiments are further conducted to investigate the contribution of adopting DUN in our PUERT.

Firstly, considering a DUN and a non-DUN with similar performances, we respectively equip them with our sampling subnet (SampNet), so as to compare the different performance gains brought by SampNet.
Concretely, we select U-Net \cite{zbontar2018fastmri} as an instance of non-DUNs, and choose our RecNet with 4 stages, named RecNet$_{s4}$, to represent DUNs.
Table~\ref{tab:dun} provides the PSNR results of U-Net with and without SampNet, as well as RecNet$_{s4}$ with and without SampNet, when ratio is 5\% and 10\% under 1D and 2D settings on Brain dataset.
It can be observed that, without SampNet, U-Net and RecNet$_{s4}$ achieve similar PSNR results on average. However, when equipped with SampNet, RecNet$_{s4}$ realizes remarkably higher PSNR increase than U-Net (1.46 dB v.s. 1.03 dB on average).
This result confirms the superiority of DUNs against non-DUNs on fully exploiting the knowledge on the learned mask.

\begin{table}[t]
\centering
\footnotesize
\setlength{\tabcolsep}{5.5pt}   
\caption{Average PSNR results for discussions on DUN when ratio is 5\% and 10\% under 1D and 2D settings on Brain dataset.}
\label{tab:dun}
\begin{tabular}{c|c|c|c|c|c}
\shline
\multirow{2}{*}{Methods} &  \multicolumn{2}{c|}{1D setting} & \multicolumn{2}{c|}{2D setting} & \multirow{2}{*}{Average} \\ \cline{2-5} 
       & 5\%    & 10\%       & 5\%      & 10\%  &  \\ \hline \hline
U-Net w/o. SampNet               &  30.03 & 33.32 & 34.66  & 37.31 & 33.83  \\
U-Net w. SampNet          &  32.09 & 33.68 & 35.65 & 38.03 & 34.86 \\
PSNR Increase            & 2.06  & 0.36  & 0.99  & 0.72  & 1.03 \\ \hline \hline
RecNet$_{s4}$ w/o. SampNet & 28.76 & 33.36 & 34.58 & 37.50 & 33.55 \\
RecNet$_{s4}$ w. SampNet  &  30.62 & 34.52 & 36.22 & 38.67 & 35.01 \\ 
PSNR Increase           & 1.86  & 1.16  & 1.64  & 1.17  & 1.46   \\ \hline \hline
PUERT                & 32.26 & 35.51 & 36.74 & 39.13 & 35.91 \\
PUERT$^{+}$          & 32.30 & 35.58 & 36.89 & 39.44 & 36.05 \\ 
PSNR Increase       & 0.04 & 0.07  & 0.15  & 0.31  & 0.14  \\ \shline
\end{tabular}
\end{table}

Secondly, in order to validate the promising superiority of DUNs on further exploiting the information in SampNet and improving the performance, we propose an advanced version of PUERT, dubbed PUERT$^{+}$, which integrates the probabilistic sampling pattern $\mathbf{P}$ in each stage of RecNet.
Specifically, for each stage, we add a module called GDM$^{\mathbf{P}}$, which is the same as GDM in Eq.~(\ref{eq:gdm}) except for replacing $\mathbf{M}$ with $\mathbf{P}$, formulated as:
\begin{equation}
\label{eq:gdm_p}
\mathbf{r}_{k}^{\mathbf{P}}=\hat{\mathbf{x}}_{k-1}-\rho_{k} \mathbf{F}^{H} \left(\mathbf{P} \odot \mathbf{F} \hat{\mathbf{x}}_{k-1}-\mathbf{y}\right).
\end{equation}
The input of each PMM is correspondingly modified to the concatenation of $\mathbf{r}_{k}^{\mathbf{P}}$ and $\mathbf{r}_{k}$, formulated as:
\begin{equation}
\mathbf{\hat{x}}_{k} =\mathbf{r}_{k}+ \mathcal{H}^{rec}_{k}(\mathcal{H}_{k}^{RB,2}({\mathcal{H}_{k}^{RB,1}}(\mathcal{H}^{ext}_{k}([\mathbf{r}_{k}, \mathbf{r}_{k}^{\mathbf{P}}])))),
\label{eq:pmm_p}
\end{equation}
where $[ \cdot ]$ is the concatenation operator.
In this way, each stage can fully use the information of not only the sampling mask $\mathbf{M}$ but also the probabilistic sampling pattern $\mathbf{P}$ from SampNet.
Compared to the \textit{hard-sampling} GDM with the binary mask $\mathbf{M}$, the newly added module GDM$^{\mathbf{P}}$ with the non-binary $\mathbf{P}$ can be regarded as a \textit{soft-sampling} version.
Table~\ref{tab:dun} provides the PSNR comparisons between PUERT and PUERT$^{+}$, when ratio is 5\% and 10\% under 1D and 2D on Brain dataset.
One can clearly see that PUERT$^{+}$ obtains consistently higher scores than PUERT across all situations, with average PSNR increased from 35.91 dB to 36.05 dB.
The 2D version achieves higher PSNR improvements than 1D, due to the utilization of more information.
Note that, taking the 2D version as an example, PUERT$^{+}$ only adds 2K learnable parameters, compared to PUERT with 404K parameters.
Above results verify the promising superiority of DUNs on further exploiting the information in SampNet and improving the performance.
More elaborate and efficient designs on further exploitation of SampNet are considered as our important future direction.

\section{Discussions and Limitations}

One weakness of PUERT is the assumption of the sampling model, in which all $k$-space locations are independent. Although, due to the control of sampling ratio, our network would learn the relative importance among $k$-space locations and obtain the dependency implicitly, we must admit that our PUERT is incapable of directly and explicitly learning the dependency among $k$-space locations.
Since such dependency is useful \cite{levine20173d} and promising to further improve the optimised sampling mask, we leave it as an important future direction.

Note that our SampNet can also be trivially generalized to other neural networks, and we consider the exploration of other elaborate architectural designs for PUERT as an important future direction. 
However, PUERT does not support sampling pattern learning for traditional model-based reconstruction methods, since we need to optimize the learnable probabilistic sampling pattern $\mathbf{P}$ in an end-to-end manner.
Considering that traditional methods still enjoy great advantages (\textit{e.g.}, fast adaptation to multiple masks and ratios, strong interpretability, and no training requirements), we consider the exploration of extending our sampling pattern learning method to traditional reconstruction methods as a future research direction.

In addition, there is still a certain distance from practical application. In actual scanning, one should also consider how to implement a specific under-sampling pattern in an MR pulse sequence, \textit{e.g.}, the constraints of hardware requirements \cite{weiss2021pilot} and the design of viable trajectories. 
And we leave the analysis on extending our PUERT to be totally applicable as an important direction for future research. 
Besides, our experiment does not consider the simulation of noise.
However, in practical applications, noise is inevitable in the measurement process \cite{gudbjartsson1995rician}, and we consider the extension of PUERT to handling noisy $k$-space data as an important future direction.

Also note that our current PUERT is restricted to the single coil CS-MRI reconstruction. However, accelerated parallel imaging \cite{pruessmann1999sense,lv2021transfer,lv2021pic} is remarkably promising to achieve higher degrees of acceleration. And we consider the combination of PUERT with multi-coil imaging as an important area of research.
There exists some literature to explore data-driven learning of sampling patterns in accelerated parallel MRI.
\cite{gozcu2019rethinking} employs a combinatorial method that lets the data decide sampling masks matched to specific parallel MRI decoders in use. 
\cite{zibetti2021alternating} proposes a learning approach that alternates between improving the sampling pattern, using bias-accelerated subset selection, and improving parameters of the variational networks. 
Insights from these studies are promising to help inspire our relevant future research on parallel imaging.

\section{Conclusions}
In this paper, we simultaneously deal with two problems in CS-MRI, \textit{i.e.}, under-sampling and reconstruction, and propose a novel end-to-end Probabilistic Under-sampling and Explicable Reconstruction neTwork, dubbed PUERT, to achieve an efficient combination of sub-sampling learning and reconstruction network training. 
Based on extensive experiments on two widely used MRI datasets, we have validated that our proposed PUERT performs favorably against state-of-the-art methods in terms of both quantitative metrics and visual quality, and achieves remarkable results even under challenging low sampling ratios. 
Besides, our detailed ablation studies confirm the importance of three components in our proposed PUERT, \textit{i.e.}, Probabilistic Under-sampling (PU), Dynamic Gradient Estimation (DGE) and Deep Unfolding Network (DUN), where we highly emphasize to adopt the DUN in PUERT so as to fully explore the information from SampNet.
In addition, an enhanced version of PUERT, dubbed PUERT$^{+}$, is also developed as an attempt to implement further exploitation of SampNet and obtain performance improvements.

{
\bibliographystyle{IEEEtran}
\bibliography{main_jstsp_arxiv}
}

\end{document}